# Thermodynamics of Fluid Polyamorphism

Mikhail A. Anisimov[1,2*], Michal Duška[1,3], Frédéric Caupin[4], Lauren E. Amrhein[1], Amanda Rosenbaum[1], and Richard J. Sadus[5]

[1]*Department of Chemical & Biomolecular Engineering and Institute for Physical Science & Technology, University of Maryland, College Park, U.S.A.*
[2]*Oil and Gas Research Institute of the Russian Academy of Sciences, Moscow, 117333, Russia*
[3]*Institute of Thermomechanics of the Czech Academy of Sciences, 182 00 Prague 8, Czech Republic*
[4]*Université de Lyon, Université Claude Bernard Lyon 1, CNRS, Institut Lumière Matière, F-69622 Villeurbanne, France*
[5]*Centre for Molecular Simulation, Swinburne University of Technology, Hawthorn, Victoria 3122, Australia*

"Fluid polyamorphism" is the existence of different condensed amorphous states in a single-component fluid. It is either found or predicted, usually at extreme conditions, for a broad group of very different substances, including helium, carbon, silicon, phosphorous, sulfur, tellurium, cerium, hydrogen and tin tetraiodide. This phenomenon is also hypothesized for metastable and deeply supercooled water, presumably located a few degrees below the experimental limit of homogeneous ice formation. We present a generic phenomenological approach to describe polyamorphism in a single-component fluid, which is completely independent of the molecular origin of the phenomenon. We show that fluid polyamorphism may occur either in the presence or the absence of fluid phase separation depending on the symmetry of the order parameter. In the latter case, it is associated with a second-order transition, such as in liquid helium or liquid sulfur. To specify the phenomenology, we consider a fluid with thermodynamic equilibrium between two distinct interconvertible states or molecular structures. A fundamental signature of this concept is the identification of the equilibrium fraction of molecules involved in each of these alternative states. However, the existence of the alternative structures may result in polyamorphic fluid phase separation only if mixing of these structures is not ideal. The two-state thermodynamics unifies all the debated scenarios of fluid polyamorphism in different areas of condensed-matter physics, with or without phase separation, and even goes beyond the phenomenon of polyamorphism by generically describing the anomalous properties of fluids exhibiting interconversion of alternative molecular states.

Subject Areas: Thermodynamics, Statistical Physics, Phase Transitions, Fluids

*To whom correspondence should be addressed. Email: anisimov@umd.edu.

# I. INTRODUCTION

"Fluid polyamorphism" is the existence of two or more amorphous condensed states in a single-component fluid [1-6]. The possibility of a liquid-liquid transition in a pure substance, in addition to ordinary vapor-liquid separation, is commonly considered as the signature of fluid polyamorphism [3, 7]. However, different amorphous phases can also exist in single-component fluids without liquid-liquid separation (first-order transition) resulting in a continuous (second-order) phase transition [8-12]. Fluid polyamorphism is found or predicted in a broad group of very different systems, including (but not limited to) helium [8, 9], sulfur [10-12], phosphorous [13], carbon [14], cerium [15], silicon [16-19], silicon dioxide [20-22], tellurium [23-25], tin tetraiodide [26, 27], and hydrogen [28-30]. Significantly, it has been also hypothesized in metastable and deeply supercooled water [31-36]. Two alternative forms of molecular arrangements are believed to exist in supercooled liquid water: a low-density structure and a high-density structure. Under certain conditions, metastable liquid-liquid separation could occur in pure water because of the existence of these two alternative structures. The hypothesized liquid-liquid metastable coexistence is not directly accessible in bulk-water experiments because it is presumably located a few degrees below the kinetic limit of homogeneous ice formation [36, 37]. Such coexistence has been reported for some atomistic water models (see review [36]), most notably in molecular simulations of the ST2 model [38]. A phase diagram similar to that predicted for water was reported for a model of supercooled silicon with the liquid-liquid transition line extending to negative pressures in the doubly metastable region [22].

The examples of polyamorphism go far beyond supercooled water or other tetrahedral fluids, such as silicon or silica. At high temperature and pressures of hundreds of GPa, highly compressed fluid hydrogen is believed to occur in two forms: atomic, metallic hydrogen and molecular, nonmetallic hydrogen [28-30]. The chemical reaction $2\mathrm{H} \rightleftarrows \mathrm{H}_2$ under these conditions may be accompanied by a first-order fluid-fluid transition. It is expected that the fluid-fluid transition line is terminated at a critical point, above which there is a gradual transformation between the two forms of highly compressed (dense plasma) hydrogen. A mixture of two interconvertible hydrogen species can be considered thermodynamically as a single-component fluid because the number of degrees of freedom is constrained by the condition of chemical-reaction equilibrium. Reversible polymerization is another example of an equilibrium chemical reaction that causes a dramatic change of substance properties. When the degree of polymerization $N$ is very large, the reaction $NA \rightleftarrows A_N$ can be considered as a second-order phase transition between the monomer phase and the solution of polymer in monomers [10]. At the second-order transition point, there is no fluid phase separation. There is no discontinuity in the density and entropy at the transition point, although there is a symmetry break.

Liquid helium and sulfur represent two well-studied examples of fluid polyamorphism without phase separation. The "lambda transition" at ~ 2 K in $^4$He, between the normal fluid and superfluid phases is a second-order transition caused by quantum Bose condensation [8]. Returning to phenomena at higher temperatures, liquid sulfur is sharply polymerized at ~ 433 K [10-12]. No fluid phase separation is observed. Phosphorous is another example of polyamorphism driven by polymerization, though is not as well-studied and, unlike polymerization of sulfur, it is claimed to be accompanied by phase separation [13].

A fundamentally important question is: what, if anything, is common to all the chemically and physically very different systems exhibiting polyamorphism? In this work, we present a generic phenomenological approach, based on the Landau theory of phase transitions [39], to describe fluid polyamorphism in a single-component substance. The approach is completely independent of the underlying molecular nature of the phenomenon. To specify this approach and calculate both phase behavior and thermodynamic properties, we consider a fluid with thermodynamic equilibrium between two competing interconvertible molecular "states" or structures. A fundamental signature of this concept is the identification of the equilibrium fraction of molecules involved in each of these alternative states.

The idea that water is a "mixture" of two different structures dates back to the 19th century [40, 41]. Rapoport used this idea to explain the high pressure melting curve maxima of some liquid metals [42]. More recently, the concept of two alternative condensed amorphous states has become a popular explanation for liquid polyamorphism in cerium caused by delocalization of Fermi electrons [15], tellurium (a competition between twofold and threefold local atomistic coordination) [23-25], tin tetraiodide (face-to-face vs. vertex-to-face orientation between the nearest molecules) [26, 27], and in water [43-49]. The variation of the relative proportion of the alternative structures with temperature and pressure, predicted in ref. [49], was used to explain the anomalous behavior of viscosity in supercooled water [50]. In a series of works by Tanaka et al., the idea of two competing liquid states was specified in terms of the alternative locally favored structures and two order parameters associated with these structures [51-54].

However, most of the previously reported versions of two-state thermodynamics considered only liquid-liquid separation and ignored vapor-liquid transition or, at best, introduced it empirically as a polynomial background part of the Gibbs energy. Hence, the complete ("global") phase diagram was not obtained. Another limitation of previous studies utilizing the two-state approach is that they all considered only polyamorphism associated with liquid-liquid separation, thus ignoring such important cases as superfluidity in helium or polymerization in sulfur. Furthermore, a broad class of systems that exhibit equilibrium interconversions of polymorphic molecules or supramolecular units, but do not exhibit polyamorphism (e.g., structural isomerization of hydrocarbons, conformation of polymer chains, folding/unfolding of protein molecules, or interconversion of stereoisomers) has not been previously unified with polyamorphic fluids.

In this work, we formulate a mean-field equation of state that globally describes both vapor-liquid and liquid-liquid transitions in the same single-component fluid. A second-order phase transition, causing fluid polyamorphism without fluid-fluid separation, is also described by this generic phenomenology. A particular variant of this global equation of state also describes the systems that do not macroscopically exhibit polyamorphism but still exhibit interconversion of polymorphic molecules. Significantly, the global equation of state is also applied for negative pressures (stretched fluids). Negative pressures are observed and studied experimentally, particularly in water [55-57], and as such they are not simply a theoretical curiosity.

We discuss two alternative mechanisms for a liquid-liquid transition in a single-component fluid. The "discrete" mechanism is driven by the existence of two distinct mixable or unmixable molecular forms or supramolecular structures. In contrast, the "continuous" mechanism, associated with isotropic two-scale nonideality in the Gibbs energy, does not require the entropy of mixing of two distinct alternative entities and the system is not constrained by the condition of interconversion equilibrium. Thermodynamically, these cases may produce similar phase diagrams and similar property anomalies, depending on interplay of the model parameters. Unambiguous discrimination of these mechanisms can be made by examining (experimentally and computationally) kinetics of structural relaxation by tuning the rate of interconversion and measuring (or simulating) the rate of relaxation of the reaction coordinate.

## II. RESULTS

### a. Generic formulation of polyamorphism in a single-component fluid

A generic thermodynamic description of fluid polyamorphism can be formulated by using the Landau theory of phase transitions [39], in which the key concept is the order parameter, a variable that characterizes the emergence of a more ordered state. The Gibbs energy (per molecule) $G$ of a single-component fluid is generally presented in the form

$$G(p,T,\phi) = G_o(p,T) + kTf(\phi) - h\phi, \qquad (1)$$

where $p$, $T$ and $k$ are the pressure, temperature, and Boltzmann's constant, respectively. In Eq. (1), $\phi$ is the order parameter. The order parameter could be either a scalar, a vector, or a tensor. The variable $h$ is a thermodynamic field conjugate to the order parameter known as the "ordering field," and $f(\phi)$ is a function whose specific form depends on the microscopic nature and symmetry of the order parameter. If the order parameter is a vector, the ordering field is also a vector. In this case the order parameter breaks the symmetry of the disordered state. We must note that Eq. (1) applies to phenomena and systems with different physical nature of the order parameter and, correspondingly, ordering field. In some cases such as magnetization (a vector), the ordering field (i.e., the magnetic field) is an independent variable, whereas generally in polyamorphic fluids the ordering field may be a function of pressure and temperature. It is also possible that some phase transitions occur only in zero ordering field because the state with non-zero field does not physically exist, e.g., the lambda transition in He$^4$ [6] or the transition from isotropic liquid to nematic liquid crystal (the order parameter is a tensor) [39]. The transition between isotropic liquid and nematic liquid crystal in a pure substance is an example of the first-order transition without phase separation, unless the order parameter (tensor of anisotropy) is coupled with the density. The existence of magnetic fluids and nematic liquid crystals makes fluid polyamorphism to be part of more general phenomena, "fluid polymorphism". In an ordinary isotropic liquid, there also could be two different types of symmetry if its molecules have two stereoisometric forms. If the liquid has different number of the stereoisomers, it will not possess a center of symmetry with respect of reflection in any plane [39].

The equilibrium value of the order parameter is found by minimizing the Gibbs energy via $\left(\partial G/\partial\phi\right)_{p,T}=0$. This minimization results in the equilibrium condition $h(p,T)=\left(\partial f/\partial\phi\right)_{p,T}$ and thus makes the equilibrium value of the order parameter, $\phi=\phi_e$, to be a function of $p$ and $T$. A particular form of $\phi=\phi_e$ depends on the nature of the order parameter. Generally, one can define $\phi_e(p,T)$ to vary between zero (alternative amorphous structure is absent) and unity (fully developed alternative amorphous structure). We emphasize that in our approach we include the ordinary vapor-liquid transition in the "background" part of the Gibbs energy $G_o(p,T)$ that is independent of $\phi$.

### b. Fluid polyamorphism induced by interconversion of molecular states

To enable the general formulation of fluid polyamorphism for the calculation of thermodynamic properties, we need to specify the nature of the order parameter and, consequently, the explicit form of the function $f(\phi)$. A unifying scenario for many, if not most, polyamorphic systems is thermodynamic equilibrium between two alternative interconvertible molecular states or supramolecular structures. This scenario is phenomenologically equivalent to "chemical-reaction" equilibrium between two alternative "species", A and B. We do not need to specify the atomistic structure of these states. They can be two different structures of the same molecule (isomers), dissociates and associates, or two alternative supramolecular structures, such as different forms of a hydrogen-bond network. Hence, the conversion of one molecular or supramolecular state to another one may not necessarily require breaking of chemical bonds.

Let $x$ be the fraction of the state B in the "chemical reaction" $A \rightleftarrows B$. This variable is also known as the "reaction coordinate" or "degree of reaction" [58]. In chemical reactions the number of atoms is conserved, while number of molecules may or may not conserved. The conservation of the number of atoms is controlled by stoichiometric coefficients. For simplicity, we first consider equal stoichiometric coefficients for A and B, meaning that the number of molecules in the reaction is conserved. Generally, the reaction $A \rightleftarrows B$ may involve different stoichiometric coefficients $\nu_A$ and $\nu_B$ (such as $2H \rightleftarrows H_2$ or, generally, $\nu_A A \rightleftarrows \nu_B B$). Specific stoichiometry may modify the relation between the reaction coordinate and the molecular fraction of each state and separate the condition of reaction equilibrium and the condition of phase equilibrium.

We specify the Gibbs energy (per molecule) given by Eq. (1) in the form

$$G = G_A + xG_{BA} + G_{mix}, \tag{2}$$

where $G_{BA} = G_B - G_A$, the difference between the Gibbs energies of the molecular entities B and A, is equivalent to $-h$, while $G_{mix}$, the Gibbs energy of mixing, is equal to $kTf(\phi)$. Furthermore, the Gibbs energy of the "background" state, $G_A$, can be identified with $G_0$, while $x$, the molecular fraction of state B, with the order parameter $\phi$.

Adopting in this section a symmetric form of the Gibbs energy of mixing, $G_{mix} = H_{mix} - TS_{mix}$ (where $H_{mix}$ is the enthalpy of mixing and $S_{mix}$ is the entropy of mixing), such that $H_{mix} - TS_{mix} = \omega x(1-x) + kTx\ln(x) + kT(1-x)\ln(1-x)$, we write [58, 59]

$$G(p,T,x) = G_A(p,T) + xG_{BA}(p,T) + kTx\ln(x) + kT(1-x)\ln(1-x) + \omega x(1-x), \tag{3}$$

where $\omega$ is the parameter of nonideality of mixing. In general, $\omega$ is a function of $T$ and $p$. The ideal-solution mixing in the Gibbs energy of mixing is represented by the ideal-gas mixing entropy $S_{mix}^{ideal} = -k\left[x\ln(x) + (1-x)\ln(1-x)\right]$. If $\omega$ does not depend on $T$, being a system-dependent constant, or depends only on $p$, the nonideality $G_{mix} - TS_{mix}^{ideal} = \omega x(1-x)$ is entirely enthalpy driven ("regular-solution mixing"). If $\omega$ is simply proportional to $T$, while being arbitrarily dependent on $p$, the nonideality is entirely entropy driven ("athermal-solution mixing"). In most real mixtures, nonideality is driven by both, enthalpy and entropy. For simplicity, in this section, we consider $\omega$ to be a constant.

The molecular fraction of state B (i.e., $x$), is a reaction coordinate. The equilibrium value of the reaction coordinate (in this particular formulation) is equivalent to the equilibrium value of the order parameter $\phi$. The chemical reaction equilibrium between A and B makes the mixture of A and B equivalent thermodynamically to a single-component fluid. Indeed, the equilibrium value of the reaction coordinate $x = x_e(T,p)$, the fraction of molecules involved in state B, is obtained from the condition of chemical reaction equilibrium [39, 58]

$$\left(\frac{\partial (G/kT)}{\partial x}\right)_{p,T} = 0, \tag{4}$$

yielding the explicit relation between the order parameter and the ordering field:

$$h = kT \ln K(p,T) = -G_{BA}(p,T) = kT \ln \frac{x}{1-x} + \omega(1-2x), \tag{5}$$

where $K(T,P)$ is the reaction equilibrium constant. In a binary mixture without chemical equilibrium, the difference between the Gibbs energies $G_{BA}$ depends on an arbitrary constant because $G_A$ and $G_B$ are independent. The chemical-equilibrium condition (4) eliminates this uncertainty, thus making $G_{BA}$ well defined.

An important practical question arises: under which experimental conditions will the system described by Eq. (3) behave either as a binary fluid mixture or as a single-component fluid? The answer depends on the separation of time scales: a system with two interconvertible fluid structures can be thermodynamically treated as a single-component fluid if the time of observation is longer than the characteristic time of reaction (fast conversion). At the opposite limit (slow conversion) the system behaves thermodynamically as a two-component mixture. In this case, the constraint imposed by Eq. (4) does not apply and the concentration of the species becomes an independent variable. Therefore, applying the "chemical-reaction" approach for the description of single-component-fluid polyamorphism assumes that the conversion is fast enough to satisfy the equilibrium condition (4) within the experimental time scale.

We emphasize that our use of the term "chemical-reaction equilibrium" does not necessarily imply that polyamorphism and liquid-liquid separation in a pure fluid involves a chemical reaction in the conventional definition, i.e., breaking of chemical bonds. Within the framework of Landau theory of polyamorphism, this terminology is phenomenologically equivalent to the condition of thermodynamic equilibrium with the Gibbs energy containing the ideal entropy of mixing of two distinct alternative states and the nonideal ("excess") Gibbs energy of mixing.

### c. Polyamorphic fluid-fluid phase separation

We note that for the symmetric Gibbs energy of mixing given by Eq. (3) the condition $\ln K(p,T) = 0$ and the condition of phase equilibrium (zero ordering field) coincide. Along the line $\ln K(p,T) = 0$, if $\omega/k_B T \leq 2$, there is only one solution of Eq. (5), that is $x = 1/2$. However, if $\omega/k_B T > 2$, this equation has two stable solutions, $x > 1/2$ and $1 - x < 1/2$. This corresponds to the coexistence of two fluid phases enriched with either A or B. This means that the line $\ln K(T,P) = 0$ is the fluid-fluid phase transition line. The temperature

$$T_c^* = \frac{\omega}{2k} \qquad (6)$$

is the critical temperature for the polyamorphic fluid-fluid transition. The critical pressure $p_c^*$, is found from the condition $\ln K(T = T_c^*, p = p_c^*) = 0$. The temperature of the fluid-fluid coexistence ("cxc") as a function of the fraction of state B is found as

$$\hat{T}_{cxc} = \frac{2(2x-1)}{\ln x(1-x)}, \qquad (7)$$

where $\hat{T}_{cxc} = T_{cxc}/T_c^*$. At the critical point $x = x_c = 1/2$. Above the critical temperature, the line $\ln K(p,T) = 0$ (a continuation of the line of phase transitions, along which $x = 1/2$) is known as the Widom line [49].

Equation (7) is equivalent to the temperature-dependence of the spontaneous (in zero field) order parameter obtained in the mean-field approximation for the Ising/lattice gas model. Indeed, introducing $M = 2x - 1$ and using $\operatorname{arctanh}(M) = \frac{1}{2} \ln \frac{1+M}{1-M}$, we obtain the well-known Ising-model mean-field result [59]:

$$M = \tanh \frac{M}{\hat{T}_{cxc}}. \qquad (8)$$

Expansion of Eq. (8) in powers of $M$ in the vicinity of the critical point yields the asymptotic power law in the mean field approximation: $M = 2x - 1 = \pm \left[ 3(T - T_c^*)/T_c^* \right]^{1/2}$.

### d. Calculation of thermodynamic properties

From Eq. (3), we can obtain

$$\frac{\partial x}{\partial p}G_{BA} + kT\frac{\partial}{\partial p}\left[x\ln x + (1-x)\ln(1-x) + \omega x(1-x)\right] = 0, \quad (9)$$

and

$$\frac{\partial x}{\partial T}G_{BA} + k\frac{\partial}{\partial T}\left[x\ln x + (1-x)\ln(1-x) + \omega x(1-x)\right] = 0. \quad (10)$$

The density and entropy are calculated from

$$\rho(p,T) = \frac{1}{V(p,T)} = \left(\frac{\partial G}{\partial p}\right)_T^{-1} = \frac{1}{V_A(p,T) - kT(\partial \ln K / \partial p)x(p,T)}, \quad (11)$$

and

$$S(p,T) = -\left(\frac{\partial G}{\partial T}\right)_p = S_A(p,T) + k(\partial \ln K / \partial T)x(p,T), \quad (12)$$

where $V_A = V_A(p,T)$ and $S_A(p,T)$ are the volume and entropy (per molecule) of state A. From the Gibbs energy, if state A and $\ln K(T,P)$ are specified we can obtain all other thermodynamic properties, such as the isothermal compressibility and heat capacity, as well as the global phase diagram that includes both vapor-liquid and liquid-liquid transitions.

e. **Specifying state A and equilibrium constant**

We have used two alternative choices of $G_A(p,T)$. One option is to adopt the chemical potential of the lattice-gas model $G_A = \mu_{lg}$. The other option is to use the chemical potential of the van der Waals fluid $G_A = \mu_{vdw}$. Both these models famously describe the transition between liquid and gaseous states and vapor-liquid coexistence. However, there is an important conceptual difference between these two models: lattice gas is a discrete model consisting of two distinct states (empty cells and occupied cells) with the entropy mathematically equivalent to the entropy of mixing in a binary fluid. The van der Waals fluid is a continuous model without distinct alternative states. In Supplemental Material (Section 1), we provide details of thermodynamic equations for the lattice-gas and van der Waals models, as well as for the fine lattice discretization model (Section 2) that uniformly describes crossover between the two alternative models [60]. The effect of the differences in these two alternative formulations of state A on the global phase diagram and properties of a polyamorphic fluid are not significant. Major effects are caused by a particular dependence of the equilibrium constant on $p$ and $T$ and by the distance from the liquid-liquid critical point to the absolute stability limit of the liquid state with respect to vapor (the liquid branch of the vapor-liquid spinodal).

The formulation on an explicit equation of state requires the specification of the equilibrium constant $K(T,p)$. A general form of the Gibbs energy change of reaction can be represented by the polynomial

$$G_{BA}(p,T) = -kT \ln K(T,p) = k\left(\lambda + \alpha p + \beta T + \gamma pT + \delta p^2 + \varepsilon T^2 + ...\right). \qquad (13)$$

Correspondingly, for the equilibrium constant:

$$-\ln K(p,T) = \frac{G_{BA}}{kT} = \frac{\lambda}{T} + \alpha\frac{p}{T} + \beta + \gamma p + \delta\frac{p^2}{T} + \varepsilon T + ..., \qquad (14)$$

where the coefficients of the polynomial represent the changes (in first approximation) of energy ($\lambda$), volume ($\alpha$), entropy ($\beta$), isobaric expansivity ($\gamma$), heat capacity ($\delta$), and isothermal compresiblity ($\varepsilon$) in the reaction $A \rightleftarrows B$. In the linear approximation,

$$G_{BA}(p,T) = k\left(\lambda + \alpha p + \beta T\right). \qquad (15)$$

In this approximation the conversion between two states is only affected by changes in energy, volume, and entropy. The phase transition line and the Widom line are defined as $\lambda + \alpha p + \beta T = 0$ with a constant slope $dp/dT = S_{BA}/V_{BA} = -\beta/\alpha$. In Supplemental Material, Section 3 we report the results for an alternative form of the equilibrium constant. These results support our conclusion on the generic character of the developed approach.

### f. Global phase diagrams and lines of extrema of thermodynamic properties

In this section we present results obtained by using $G_A(p,T)$ for the chemical potential of the lattice-gas model. Essentially similar results, presented in Supplemental Material, Section 4, are obtained when adopting $G_A(p,T)$ for the chemical potential of the van der Waals model.

A typical phase diagram, calculated from Eq. (3) with $G_{BA}(p,T)$ using the linear approximation given by Eq. (15), for the polyamorphic lattice-gas model is presented in figures 1a and 1b. Dimensionless values of temperature ($\hat{T}$) and pressure ($\hat{p}$) are relative to the critical parameters of the vapor-liquid critical point (CP1). The parameters of the model for this particular case are, $\lambda/kT_{c1}=0.5$, $\alpha/\rho_{c1}kT_{c1}=-0.05$, $\beta/k=-1.5$, and $\omega/kT_{c1}=0.6$. Figures 1a and 1b show the vapor-liquid and liquid-liquid coexistence (with the liquid-liquid critical point, CP2, as a simple example located at a positive pressure for the selected parameters), the absolute stability limit of the liquid state with respect to vapor, and the Widom line. The right branch of the vapor-liquid spinodal, which is the absolute stability limit of liquid with respect to vapor (obtained as the locus of maxima of the isobars), demonstrates re-entrant behavior at the densities close to the liquid-liquid critical density.

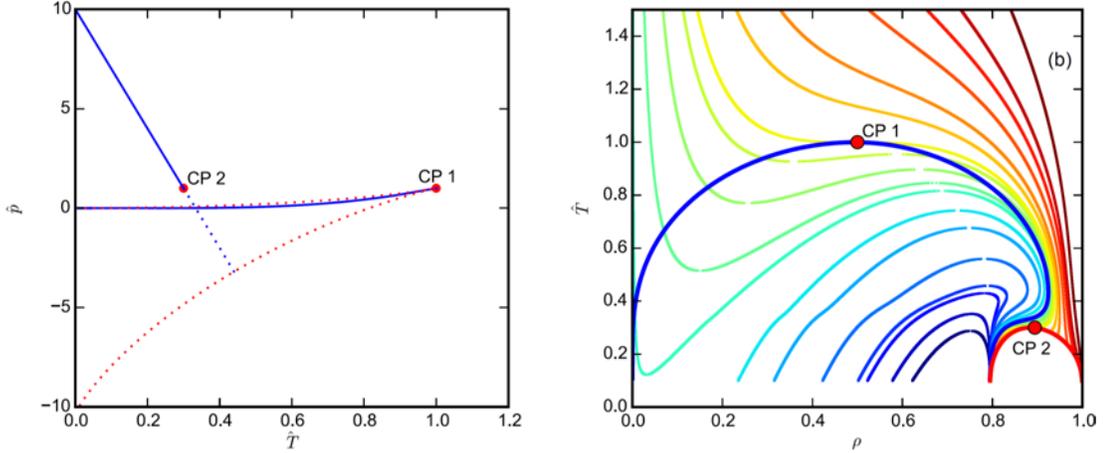

**Figure 1.** Phase diagram for a polyamorphic lattice gas; $\hat{T}=T/T_{c1}$ and $\hat{p}=p/p_{c1}$. (a) Pressure/temperature diagram. The blue curves are either vapor-liquid or liquid-liquid transitions; CP1 and CP2 are the vapor-liquid and liquid-liquid critical points assigned (as an example) to be at the same isobar. The red dotted curves are the liquid and vapor branches of liquid-vapor spinodal) and the blue dotted line is the Widom line. (b) Temperature/density diagram. The thick blue and thick red curves are the vapor-liquid and liquid-liquid coexistence, respectively. The multicolor curves are selected isobars.

Figure 2 demonstrates an example of the behavior of the isothermal compressibility along three selected isobars: above, at and below the critical pressure. For the above liquid-liquid critical pressure case (green), the pressure for both CP1 and CP2 are assigned to be equal. One can notice the divergence of the isothermal compressibility at the critical points and at the vapor-liquid spinodal.

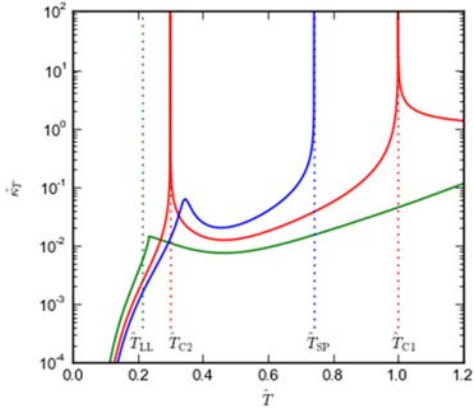

**Figure 2**. The dimensionless isothermal compressibility $\hat{\kappa}_T = \rho^{-1}\left(\partial\rho/\partial\hat{p}\right)$ along three selected isobars: above the liquid-liquid critical pressure, assigned equal to the vapor-liquid critical pressure, (green), at the critical pressure (red), and below the critical pressure (blue). $\hat{T}_{c1} = 1$ and $\hat{T}_{c2} = T_{c2}/T_{c1}$ are the vapor-liquid and liquid-liquid critical temperatures, $\hat{T}_{LL} = T_{LL}/T_{c1}$ is the temperature of the liquid-liquid transition at the selected isobar.

The location of the liquid-liquid critical point depends on the interplay of two essential parameters, $\lambda$ (the energy change of reaction at zero pressure) and $\omega = 2kT_{c2}$ (the nonideality of mixing of the alternative states). As shown in Figure 3, by tuning $T_{c2} = \omega/2k$ from zero to a certain positive value (depending on $\lambda$) the model evolves from a "singularity-free" scenario ($T_{c2} = 0$) to a "critical-point-free" scenario (the liquid-liquid critical point is located beyond the absolute stability limit of liquid state with respect to vapor).

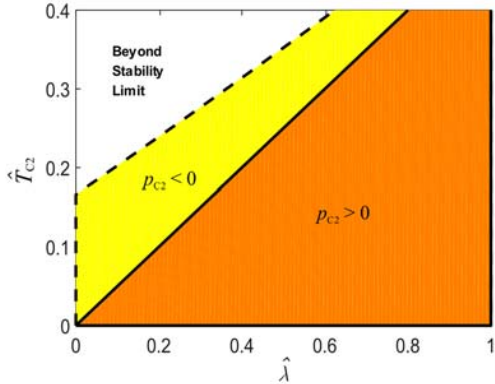

**Figure 3.** Parameterized phase diagram for polyamorphic lattice gas in terms of nonideality of mixing of the alternative states A and B, $\hat{T}_{c2} = \omega/2kT_{c1}$, and the energy difference of A and B, $\hat{\lambda} = \lambda/kT_{c1}$. The volume difference of states A and B is taken constant. The singularity-free scenario corresponds to $\hat{T}_{c2} = 0$. The critical point-free scenario is favored by stronger nonideality and smaller energy difference.

Stokely et al. [61] studied the effects of hydrogen bond cooperativity on the behavior of supercooled water. The authors introduced two major parameters: the strength of the directional component of the hydrogen bond and the degree of hydrogen bond cooperativity. If the degree of hydrogen bond cooperativity is zero, the neighboring bonds are formed independently. We note that if the strength of the directional component of the hydrogen bond is identified with the energy change of reaction ($\lambda$) and the degree of the cooperative component of the hydrogen bond is identified with the nonideality parameter $\omega \propto T_{c2}$ then the phase diagram presented in Figure 3 is essentially similar to that obtained by Stokely et al. [61]

Tuning the distance of the liquid-liquid critical point from the absolute stability limit of the liquid state with respect to vapor results in dramatic change in thermodynamic behavior of the system and, especially, in the pattern of extrema in thermodynamic properties [62, 63]. In particular, the locus of density maximum/minimum is one of the most characteristic features of polyamorphic liquids. The salient points of this locus are interrelated through thermodynamic relations, with the extrema loci of thermodynamic response functions, such as the isothermal compressibility along isobars or the isobaric heat capacity along isotherms [63, 64]. Furthermore, since the extrema loci are experimentally observed for a broad range of temperatures and pressures, including thermodynamically stable regions, their shape provides important information for modelling liquid polyamorphism, especially if the liquid-liquid transition is experimentally inaccessible [65].

The evolution of the extrema loci upon tuning the location of the liquid-liquid critical point is demonstrated in Figure 4 from (a) ("singularity-free" scenario) to (d) ("critical-point-free" scenario). The pattern of extrema loci in Figure 4a demonstrates a singularity-free scenario [62] which is relevant to those tetrahedral systems that do not exhibit a metastable liquid-liquid separation, such as the mW model of water [66], but still exhibit interconversion between alternative states. The pattern presented in Figure 4b (a "regular polyamorphism" scenario, the liquid-liquid critical point is located at a positive pressure) is observed in the popular ST2 [67] and TIP4P/2005 [68] atomistic water models. The additional (shallow) extrema of the heat capacity, observed in this case, is unrelated to the liquid-liquid transition and is specific to the model adopted for state A. The extrema are also unrelated to the so-called "weak" extrema of the heat capacity and isothermal compressibility reported by Mazza et al. [69] that emanate from the liquid-liquid critical point and which are specific for their "many-body model" of water. The case presented in Figure 4c is a degenerate one as the critical point coincides with the vapor-liquid spinodal. Finally, Figure 4d presents the case in which the transition line remains of first-order until the liquid becomes unstable with respect to vapor ("critical-point-free" scenario [70]). We note that the critical-point-free scenario is a variant of the "stability limit conjecture" proposed by Speedy [71]. Speedy viewed the cause of the anomalies of water in a continuous instability line which "bounds the metastable superheated, stretched, and supercooled states". In the critical-point-free scenario, this instability line is realized by the union of the liquid- liquid (present in our model but not shown for clarity) and liquid-vapor spinodal. One can notice that the vapor-liquid spinodal in Figure 4 remains continuous and smooth even when it intercepts the liquid-liquid transition line (Figure 4d). This is not generic, being a result of the simple linear form of $G_{BA}(T,p)$, given by Eq. (15), that was used for the calculations. This form implies that the compressibilities of states A and B are the same.

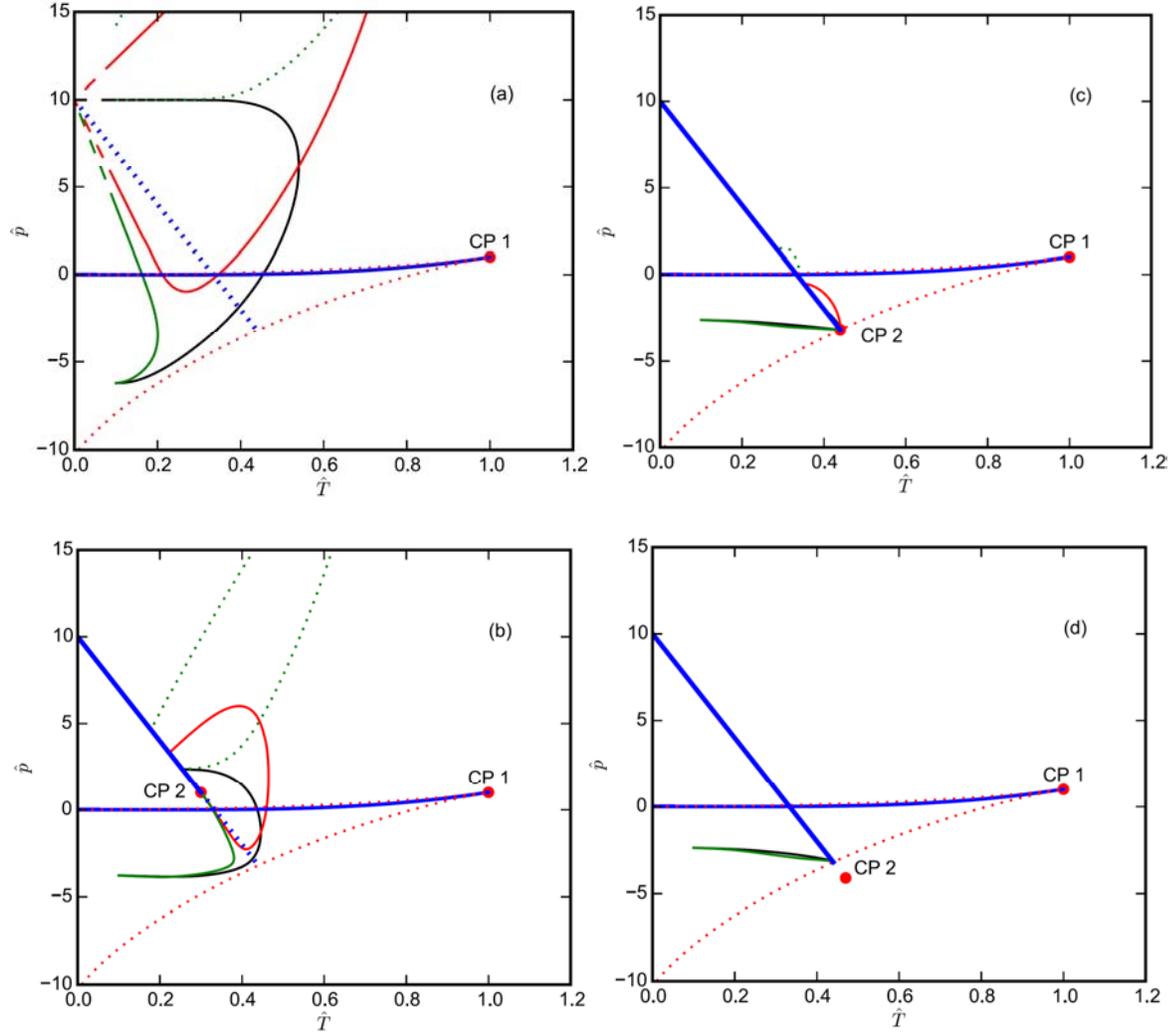

**Figure 4.** Evolution of the pattern of the extrema loci upon tuning the location of the liquid-liquid critical point. Black is the density maximum or minimum; red is the isothermal compressibility maximum or minimum along isobars, green is the isobaric heat capacity maximum or minimum along isotherms; dotted green shows additional (shallow) extrema of the heat capacity unrelated to the liquid-liquid transition; dotted red are two branched of the liquid-vapor spinodal; blue dashed is the Widom line; red dots are the vapor-liquid (CP1) is the liquid-liquid critical point (CP2). (a) a "singularity-free" scenario – the critical point is at zero temperature, thus it is not labeled as CP2; extrapolations of the extrema loci to zero temperature are shown as dashed lines; (b) a "regular" scenario – the critical point CP2 is at a positive pressure; (c) the critical point coincides with the absolute stability limit of the liquid state; (d) a "critical point-free" scenario – the "virtual" critical point CP2 is located in the unstable region.

The most dramatic result of the evolution of the extrema loci is the shrinking and eventual disappearance of the maximum density locus upon the transition from the singularity-free scenario to the critical-point-free scenario. This effect is observed for both choices of state A, the lattice gas and van der Waals models, with various sets of the model parameters (see Supplemental Material, figures S9 and S10) and has been recently observed in models of doubly metastable silicon [72] and silica [73]. To investigate in what degree this effect is common, it would be worth examining other models for state A, which could be both more realistic and specific to different polyamorphic systems or models.

Another remarkable peculiarity of liquid polyamorphism, which has not been reported previously in the literature, is a singularity ("bird's beak") in the liquid-liquid coexistence curve when the critical point coincides with the liquid-vapor spinodal (Figure 4c). This singularity is associated with the common tangent of the liquid-liquid coexistence and vapor-liquid spinodal in the $p-\rho$ and $T-\rho$ diagrams as shown in Figure 5 (a) and (b). It is also observed for the van der Waals choice for state A (see Supplemental Material, figure S11). We believe this effect is a thermodynamic requirement. A similar shape for fluid-fluid coexistence is observed in dilute binary solutions near the vapor-liquid critical point of the pure solvent when two spinodals have a common tangent in the $p-\rho$ and $T-\rho$ diagrams [74].

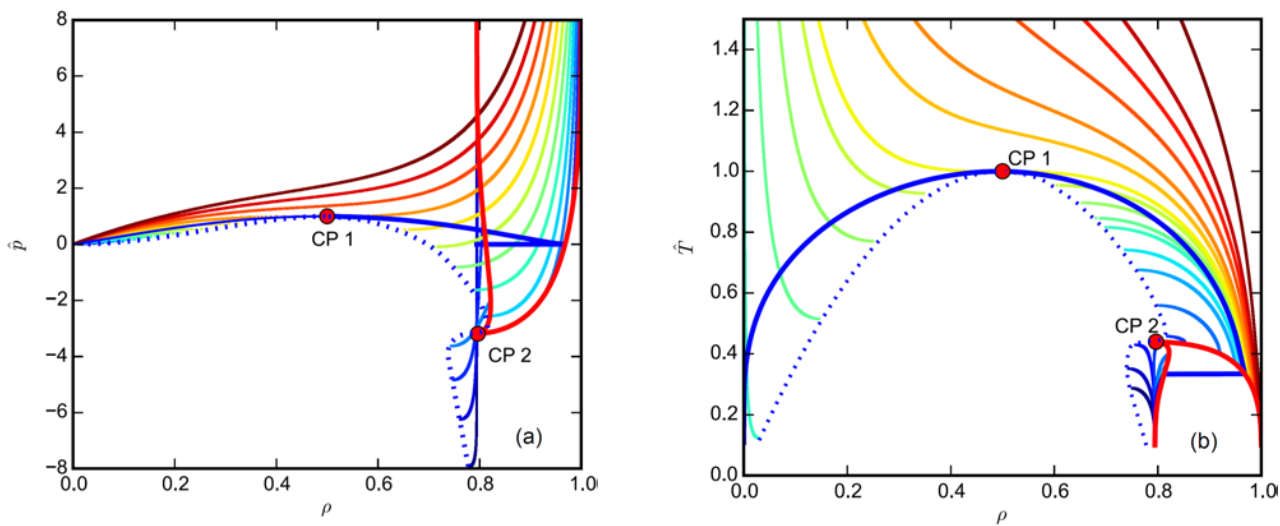

**Figure 5.** A singularity ("bird's beak") in the liquid-liquid coexistence curve if the critical point coincides with the liquid-vapor spinodal. (a) $p-\rho$ diagram, (b) $T-\rho$ diagram. Thick blue curve is the vapor-liquid coexistence; thick red is the liquid-liquid coexistence; CP1 and CP2 are the vapor-liquid and liquid-liquid critical points, respectively; dotted blue is the liquid-vapor spinodal; multicolor curves are selected isotherms (a) and selected isobars (b).

### g. Fluid polyamorphism without or with phase separation: superfluid transitions in liquid $^4$He and $^4$He-$^3$He mixtures

The second-order phase transitions ("lambda transitions") of superfluidity in pure $^4$He and $^3$He helium isotopes are arguably the most famous examples of liquid polyamorphism without phase separation [8, 9]. The formation of the superfluid is associated with the formation of a Bose-Einstein condensate. In $^4$He, the lambda transition between the normal fluid and superfluid phases occurs at ~ 2 K [8], while $^3$He forms a superfluid phase (A or B, depending on pressure) at a temperature below 0.0025 K [9].

In the mean-field approximation, polyamorphism in helium-4 is described near the transition temperature $T_\lambda(p)$ by Eq. (1) with $h(p,T) = |\mathbf{h}| = 0$, $\phi(p,T) = |\mathbf{\psi}|$ (a two-component vector order parameter, the wave function in the theory of Bose-Einstein condensation [6,7], containing real and imaginary parts) and $f(\phi)$ given by a Landau expansion [37]:

$$f(T,p) = \frac{1}{2} \frac{T - T_\lambda(p)}{T_\lambda(p)} |\mathbf{\psi}|^2 + \frac{1}{4} u |\mathbf{\psi}|^4, \tag{16}$$

where $u$ is a coupling constant. The superfluid phase below $T_\lambda(p)$ can be viewed phenomenologically as a two-state "mixture" with the fraction of the superfluid component controlled by thermodynamic equilibrium. The order parameter in the mean-field approximation changes with temperature as $|\mathbf{\psi}| = \pm\left[(1/u)(T_\lambda - T)/T_\lambda\right]^{1/2}$. In contrast, the experimentally observable physical property, the fraction ("density") of the superfluid component, is a scalar, changing along isobars as $\rho_{sf} \propto |\mathbf{\psi}|^2 \propto T_\lambda(p) - T$. The transition is continuous, occurring in pure helium-4 without phase separation. However, liquid-liquid phase separation is observed in a mixture of $^4$He and $^3$He, where the lambda transition at the tricritical-point concentration of $^3$He becomes a first-order transition [75]. The physical origin of tricriticality in the $^4$He-$^3$He mixture is a coupling between the vector order parameter $\psi$ and concentration $c$ (a scalar) of $^3$He [75, 76]. The function $f(|\mathbf{\psi}|, c)$ will contain an invariant $\propto |\mathbf{\psi}|^2 c$. In contrast to the ordinary parabolic fluid-fluid coexistence, the mean-field shape of the tricritical liquid-liquid coexistence is angle-like: the difference of concentration of the mixture and its tricritical value is a linear function of temperature, $c - c_{\text{tct}} \propto \rho_{sf} \propto T_{\text{tct}} - T$. Another remarkable feature of tricriticality is that in three dimensions it is essentially a mean-field phenomenon (with small logarithmic corrections) [75], thus making Landau theory a valid approximation.

The Landau theory, applied to superfluidity, implies that fluid polyamorphism without phase separation (a second-order transition) is associated with a vector order parameter. If the order parameter is a tensor (isotropic-nematic transitions) the transition between two fluid phases will be first-order, but, nevertheless, not necessarily with phase separation [76]. Phase separation will only emerge if the vector (or tensor) order parameter is coupled with a scalar (density or concentration).

The two-state interpretation of superfluidity certainly does not imply that there is a chemical-reaction equilibrium between alternative two states in helium. However, there is a remarkable analogy that underlines the common two-state phenomenology of polyamorphic fluids. In the two-fluid superfluidity model, the superfluid state has zero entropy. The total entropy is due to the normal fluid, and can be calculated by using Bose statistics and the excitation spectrum of helium [8]. The next section demonstrates how a similar asymmetric entropy emerges in the Gibbs energy of mixing for a two-state fluid that undergoes an equilibrium reaction of polymerization. Analogously to helium, In the infinite-degree polymerization limit, the contribution from the polymer chain to the entropy of mixing vanishes.

### h. Fluid polyamorphism caused by polymerization without or with phase separation

The transition to polymeric liquid sulfur at a temperature about ~ 433 K is another example of fluid polyamorphism without phase separation. The properties of sulfur near the polymerization transition are completely reversible as is the case for a continuous phase transition. Using a Heisenberg *n*-vector model (*n* is the number of vector's components) in the limit $n \to 0$, Wheeler et al. [77, 78] explained the polymerization in sulfur as a second-order phase transition in a weak external field. An earlier theory by Tobolsky and Eiseberg [10] describes the temperature dependence of the extent of polymerization in terms of a second-order phase transition in the mean-field approximation. The situation for real sulfur is more complicated because the polymerization of sulfur into its supramolecular structure occurs upon heating [10, 11], since liquid sulfur contains octamers that are to be broken to undergo polymerization. Furthermore, according to Dudowicz et al. [79] polymerization in actin is strictly equivalent to a thermodynamic phase transition only in the limit of zero concentration of the initiator.

Here we consider the simplest scenario, namely an equilibrium reaction of polymerization $N\text{A} \rightleftarrows \text{A}_N$ in the liquid phase of monomers A. In the limit $N \to \infty$, this reaction is equivalent to a second-order phase transition in zero field between the phase containing only monomers (state A) and the phase containing a solution of the infinite polymer chain in the monomers (state B). The phenomenon is equivalent to a second-order transition because the volume fraction of polymer is continuous at the starting point of polymerization, while its derivative is discontinuous. For polymerization in an incompressible liquid system, the volume fraction of polymer is proportional to the fraction of polymerized monomers, $x$. If the solvent molecules are just nonpolymerized monomers, this transition is described thermodynamically by the Flory mean-field theory of polymer solutions [80, 81] constrained by the equilibrium condition of polymerization. In the Flory theory, the Gibbs energy per monomer

$$G(p,T,x) = G_\text{A} + xG_\text{BA} + kT\frac{x}{N}\ln x + kT(1-x)\ln(1-x) + \omega x(1-x). \tag{17}$$

In the simplest approximation, the interaction parameter $\omega$ can be assumed to be independent of temperature, $\omega = \frac{k\Theta}{2}$, where $\Theta$ is a temperature of phase separation in the limit $N \to \infty$, $x \to 0$ (the "theta temperature"). At temperatures much higher than the theta temperature (when the interaction parameter is negligible), the infinite chain exhibits a self-avoiding walk in solution of monomers [82].

For a reversible reaction at the condition $\omega \ll kT$, the enthalpy of mixing can be neglected and the chemical-reaction equilibrium condition reads

$$G_\text{BA}(p,T) = -TS_\text{BA} = \frac{kT}{N} - kT + \frac{kT}{N}\ln x - kT\ln(1-x). \tag{18}$$

Specifying (just for simplicity) the Gibbs energy change of reaction as $G_\text{BA}(p,T) = k(\lambda + \alpha p + \beta T)$, we obtain the temperature dependence of the polymer volume fraction along isobars, presented in Figure 6. At a finite degree of polymerization, there is no sharp transition between the monomer-reach and polymer-reach states. This case corresponds to a "singularity-free" scenario in the two-state thermodynamics (there is no polyamorphism, but there is interconversion), although the asymmetry (with respect to the Widom line) in the equilibrium fraction of polymerized molecules is very strong at large $N$. However, in the limit $N\to\infty$, $G_{BA}(p,T)=-TS_{BA}=-kT-kT\ln(1-x)$, the polymer volume fraction is zero at all temperatures above the transition temperature, $T=T_\lambda(p)$, defined by the condition

$$-kT\ln(1-x)=\lambda+\alpha p+\beta T+kT=0. \tag{19}$$

We note that in this highly asymmetric case the condition $\ln K=\lambda+\alpha p+\beta T=0$ and the condition of phase equilibrium at the transition temperature, $\lambda+\alpha p+\beta T+kT=0$ are not the same. One can also notice that near the transition temperature the polymer volume fraction changes linearly as a function of $T-T_\lambda(p)$, suggesting that, like in the case of superfluidity in helium, the actual order parameter for polymerization in the limit $N\to\infty$ is proportional to $x^{1/2}$.

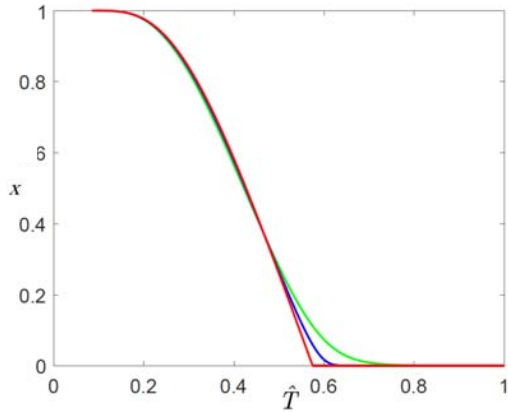

**Figure 6.** Fraction of polymerized monomers as a function of temperature for different degrees of polymerization. Green curve corresponds to $N=10$, blue $N=30$, and red $N\to\infty$. The polymerization in the limit $N\to\infty$ is equivalent to a continuous (second-order) phase transition in zero field.

This is polyamorphism without phase separation, purely driven by the extraordinary asymmetry in the entropy of mixing in the limit $N\to\infty$. Therefore, in this (asymmetric entropy-driven, no heat of mixing) case, the behavior of the system is fundamentally different from the case of nonideal mixing-driven polyamorphism with phase separation, when $N$ and $\omega$ are finite.

We emphasize that the meaning of the order parameter for the system in which two interconvertible states are controlled by chemical-reaction equilibrium changes from a scalar for all finite $N$ (in the purely symmetric case, $N=1$, this is a fraction of conversion, $x$) to a specific (zero-component vector) order parameter $\psi$, with $x\propto|\psi|^2\propto T_\lambda(p)-T$, associated with self-avoiding walk singularities of the infinitely long ($N \to \infty$) polymer chain [81-83]. This makes the infinite-degree of polymerization, at least in the mean-field approximation, to be analogous to the two-state model of superfluidity. The sharpness of polymerization fraction with the temperature variation, $\delta T / T$, the parameter that controls crossover between the singularity-free scenario (finite $N$ and zero $\omega$) and polyamorphism with a lambda transition (infinite $N$ and zero $\omega$), is $\delta T / T \propto N^{-1/2}$ [82, 83]. If $N$ is finite, the possibility of polyamorphism is always associated with phase separation and requires the existence of nonideality in the Gibbs energy of mixing (finite interaction parameter $\omega$). We note, however, that polyamorphic liquid-liquid separation (finite $N$) could, in principle, be entropically driven if $\omega$ is simply proportional to $T$, while being dependent on $p$ [49].

In addition to pure sulfur, Wheeler [84] considered polymerization of sulfur in a molecular solvent. If the mutual attraction between the monomers fragments of the polymer chain is stronger than between the chain fragments and solvent molecules, at a certain temperature, equivalent to the theta temperature, the transition to the polymer-rich phase could be accompanied by phase separation in the solution. The line of second-order phase transitions becomes the line of first-order transition at a tricritical point [39, 75, 82]. Therefore, the theta point of the infinite polymer chain in the solution of small molecules is equivalent to a tricritical point. Within the framework of Landau theory, one can interpret the emergence of the tricritical point as a result of coupling between the polymerization order parameter $\psi$ and concentration.

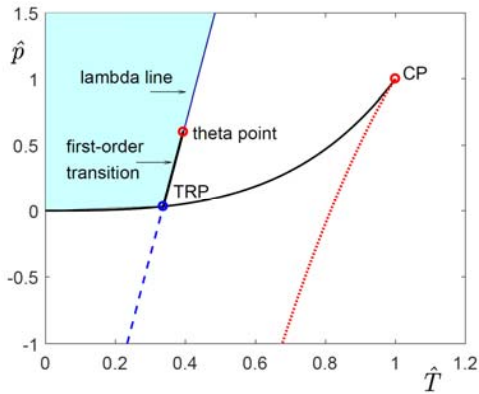

**Figure 7.** Generalized phase diagram of a fluid exhibiting equilibrium polymerization into an infinite chain. Blue area is the polymeric phase; Thick black line and curve are first-order phase transitions (coexistence between monomer and polymer phases and between vapor and liquid, respectively); CP is the vapor-liquid critical point; the theta point is equivalent to a tricritical point which separates the second-order and first order phase transitions; TRP is the triple point (monomers, polymer enriched phase, and vapor coexist). Dotted red curve represents the absolute stability limit of liquid with respect to vapor. Blue dashed line is continuation of the liquid-liquid transition line into the metastable region.

A similar phenomenon could, in principle, exist in a pure polymerizing molecular liquid if the attraction between the monomers fragments of the emerging polymer chain is stronger than between the chain fragments and nonpolymerized monomers. Then the monomers and the monomer solution enriched with the polymer will separate below the theta temperature (see figure S12 in Supplemental Material). The emergence of liquid-liquid separation and tricriticality in a single-component polymerizing fluid requires a strong coupling between polymerization and density, through the interaction term $\omega x^2 \propto |\psi|^2 x$. A generalized phase diagram of a fluid exhibiting infinite-chain polymerization with phase separation below the tricritical point is presented in Figure 7. In reality, phase separation is rare because it requires significant nonideality in interactions between the fragments of the polymer chain and its monomers. At high pressures, a sulfur melt undergoes a nonmetal-metal first-order transition [85], similar to that earlier found in selenium [86]. However, this transition is unrelated to polymerization in sulfur at atmospheric pressure, which occurs as a second-order transition. Polymerization in phosphorus, unlike polymerization in sulfur, is accompanied by phase separation [13]. If the degree of polymerization in phosphorus could be infinite (practically, $\delta T / T \propto N^{-1/2} \ll 1$), the liquid-liquid coexistence line would be separated from second-order transitions by a tricritical point, i.e., not an ordinary critical point. This thermodynamic requirement brings interesting questions to the interpretation of polyamorphism in phosphorous, given by Yarger and Wolf [87], as being associated with ordinary liquid-liquid criticality. This interpretation is unambiguously correct only if the parameter $\delta T / T \propto N^{-1/2}$ is not very small. For finite, but small $\delta T / T \propto N^{-1/2}$, a crossover between, ordinary criticality and tricriticality ($\delta T / T \propto N^{-1/2} \to 0$) should be taken into account. However, such interpretation seems to be indeed adequate for the polyamorphic behavior of triphenyl phosphite [88].

i. **Liquid-liquid transition in a single component fluid without interconversion of discrete molecular states**

In a single-component fluid, the existence of a liquid-liquid separation, in addition to the vapor-liquid transition, does not necessarily require the existence of distinct interconvertible molecules or molecular structures. Indeed, the Landau theory of phase transitions can phenomenologically describe this scenario without any reference to such interconversion.

Let the Gibbs energy of a fluid be described by Eq. (1) with the ordering field $h(p,T)$. The ordinary vapor-liquid transition is described by $G_o(p,T)$. The origin of a possible liquid-liquid transition and the nature of the order parameter depends on a particular form of the function $f(\phi)$. If we adopt a continuous free-energy model for this function, e.g. in the van der Waals-like form

$$f(\phi) = \phi \ln \frac{\phi}{1-\tilde{b}\phi} - \frac{\tilde{a}\phi^2}{kT}, \tag{19}$$

the derivative of this function yields the expression for the ordering field in the form of the chemical potential of a van der Waals fluid:

$$h = \frac{\partial f}{\partial \phi} = kT\left(\frac{1}{1-\tilde{b}\phi} + \ln\frac{\phi}{1-\tilde{b}\phi}\right) - 2\tilde{a}\phi, \tag{20}$$

where the interaction parameter $\tilde{a}$ defines the second energy scale and $\tilde{b}$ is the second distance scale. From Eq. (20), using a particular form of $h = h(T,p)$, e. g., $h = \lambda + \alpha P + \beta T$, one can obtain the equilibrium value of the order parameter $\phi_e = \phi(p,T)$. The fundamental difference between the continuous and reaction-equilibrium approaches is the definition of the ordering field $h = h(T,p)$. In the reaction-equilibrium approach the pressure/temperature dependence of $h$ is controlled by the condition of reaction equilibrium. In the continuous approach $h$ is just a specific part of the chemical potential; its particular pressure/temperature dependence is to be determined by the condition of liquid-liquid equilibrium ( $h = 0$ ). Only in the case, when the function $f$ has a form with the symmetric entropy, like in the lattice-gas model, the definitions of the ordering field in continuous and discrete (if it is also symmetric) approaches are equivalent.

We note that in the continuous scenario the order parameter is not associated with an equilibrium fraction of molecules involved in a certain state because there is no entropy of mixing of two distinct species in the function $f(\phi)$. Instead, the order parameter originates from the additional energy and length scales in the intermolecular potential, being phenomenologically associated (in first approximation) with the excess entropy $S_{ex} = S - S_0 = -\beta\phi$ and excess volume $V_0 = \partial G_0 / \partial p$, where $V_0 = \partial G_0 / \partial p$ and $S_0 = -\partial G_0 / \partial T$. The order parameter is zero for a simple fluid (which is described by a one-scale Gibbs energy) and changes from zero to unity as a function of $p$ and $T$.

In particular, for the given (van der Waals-like) example, the critical value of the order parameter $\phi_c = 1/3\tilde{b}$ and the critical temperature $T_c^* = 8\tilde{a}/27k\tilde{b}$. However, in the vicinity of the critical point of phase separation the function $f(\phi)$ can be symmetrized by an appropriate redefinition of the order parameter and the ordering field [89]. The discrete scenario can also be asymmetric, due to either asymmetric entropy of mixing or asymmetric heat of mixing. Generally, for such cases, the order parameter is not just a fraction of conversion. It will be defined through a coupling between the reaction coordinate (fraction of conversion), density, and entropy, being a combination of all these variables.

The possibility of fluid polyamorphism without interconversion of discrete molecular states is limited to liquid-liquid separation and deemphasizes structural difference between the alternative liquid phases. In particular, this scenario excludes polyamorphism without phase separation caused by infinite-degree polymerization (sulfur) and liquid-liquid separation accompanied by chemical reaction (phosphorus, hydrogen). It also excludes liquid-liquid separation of interconvertible stereoisomers and self-assembly (see Supplemental Material, Section 8).

## III. DISCUSSION

### a. Is the "critical-point-free" scenario realistic for supercooled water?

One result, reported in section **f**, have practical implications. There is an ongoing discussion in the scientific community on the possibility of a "critical-point-free" scenario in silicon, silica, and supercooled water, if the first-order liquid-liquid transition line could continue into the stretched liquid state (doubly metastable) crossing the vapor-liquid spinodal [70-73, 90, 91]. This scenario is illustrated in Figure 4d. In this scenario the locus of density maxima disappears, collapsing into the transition line at negative pressures. In contrast, the locus of density maxima for real water is observed experimentally at positive pressures. This phenomenon of shrinking the density maximum line is reproduced for both the van der Waals and lattice gas models for state A and for different forms of $G_{\text{BA}}(p,T)$ (see the Supplemental Material, figures S9d and S10). In fact, we tried many different combinations of the parameters in the two-state model and always found the same behavior. Moreover, the same collapse has been recently observed in a doubly metastable models of silicon [70 and silica [71]. Shrinking of the density maxima locus in a regular critical-point scenario with respect to a singularity-free scenario, similar to that seen in Figures 4a and 4b, was also observed by Truskett et al. [92] in an associating fluid model with directional interactions.

However, we must note that for one alternative model (the modified van der Waals model of Poole et al. [90]) , even in the case of the critical-point-free scenario, the density maximum line still exists at positive pressure. Therefore, the mere existence of the density maximum line in real water cannot reject the critical-point-free scenario. It could be possible that shrinking the density maximum locus would not always result in its disappearance. Experiments in the doubly metastable region can resolve this problem. Recent experiments on water at negative pressure [63] have observed a maximum in isothermal compressibility along isobars. This makes a strong case in favor of the second-critical-point or singularity-free scenarios. These two scenarios require the existence of a compressibility maximum at negative pressures (Figure 4), whereas the critical-point-free scenario predicts the divergence of the compressibility at the liquid-liquid spinodal (that is crossed upon cooling at negative pressures in this scenario).

### b. Is Landau theory sufficient to unify different polyamorphic phenomena?

In this work we argue that the phenomenon of fluid polyamorphism can be unified by the Landau theory of phase transitions. Landau theory is a mean-field approximation that neglects the effects of fluctuations on thermodynamic properties [39, 76, 93]. However, these effects are dominant only in the immediate vicinity of the fluid-fluid critical points and second-order phase transitions and they do not qualitatively change the phase diagrams. Furthermore, the effects of fluctuations are insignificant for first-order transitions and near tricritical points [39, 93]. Effects of fluctuations can be incorporated into the two-state thermodynamics through a well-developed crossover procedure by renormalizing the function $f(\phi)$ in Eq. (1), as described in ref. [94]. In other words, Landau theory is sufficient to address all basic issues of polyamorphic fluid phase behavior. The concept of symmetry breaking at the transition point is more important. Depending on the symmetry of the order parameter, fluid polyamorphism may or may not be accompanied by phase separation. If the order parameter is a scalar, the first-order transition between fluid phases may be terminated by a critical point. If the order parameter is a vector, a second-order transition without phase separation is possible. Moreover, coupling between scalar and vector order parameters could cause tricriticality and first-order transition in the system that otherwise would demonstrate a second-order polyamorphic transition.

### c. Can one discriminate, experimentally or computationally, between "discrete" and "continuous" approaches to fluid polyamorphism?

While the symmetry of the order parameter (scalar vs. vector) can be elucidated by the study of polyamorphic phase behavior, discrimination between two alternative approaches (continuous vs. discrete) to fluid polyamorphism in the systems with a scalar order parameter and without an obvious molecular interconversion, is a more delicate task. For the description of liquid-liquid transitions without well-defined discrete molecular states, the difference between these approaches is somewhat similar to that between the descriptions of vapor-liquid transition either by the lattice-gas model or van der Waals model (see Supplemental Material, Sections 1 and 2). In the continuous case the function $f(\phi)$ in Eq. (1) does not contain the entropy of mixing of two alternative states. Instead, this function may have a form similar to the asymmetric van der Waals-like free energy. However, the difference between the vapor-liquid transitions in the symmetric lattice-gas model and asymmetric van der Waals model is subtle. Moghaddam et al. [60] developed a "fine lattice discretization" crossover procedure that uniformly describes these two models (see Supplemental Material, Section 2). Similarly, the alternative formulations of the origin of liquid-liquid separation in a pure fluid, namely the existence of two interconvertible states or the existence of additional interaction energy and distance scales in an isotropic intermolecular potential, may generate very similar phase diagrams. Furthermore, both approaches, discrete and continuous, may generate similar extrema lines in the singularity-free scenario $(T_{c2} = 0)$. For example, Poole et al. [90] and Truskett et al. [92] proposed an extension of the van der Waals equation that incorporates the effects of the network of hydrogen bonds that exist in liquid water. They did not use the concept of the reaction equilibrium between the two alternative structures, although a possible relation between their models and two-state thermodynamics has not yet been investigated.

The question arises: can these alternative approaches be discriminated either experimentally or computationally? The discrete approach is obviously required for the description of polyamorphism caused by a well-defined chemical reaction (hydrogen, sulfur, phosphorus) or interconversion of polymorphic molecules. Depending on the stoichiometric coefficients, the entropy of mixing may or may not have a symmetric form. A hypothetical example of a discrete approach with the perfectly symmetric entropy of mixing is equilibrium folding-unfolding of a single molecule. If the conformers of this molecule do not attract each other, at certain temperature they may separate. A similar case is the symmetric phase separation of stereoisomers. A recent simulation study [94] has demonstrated the possibility of spontaneous chiral symmetry breaking in a single-component racemic (achiral) fluid upon cooling through a critical point of liquid-liquid separation. However, for other debated examples of polyamorphism, including metastable liquid water, the question of the existence and interconversion of two discrete states can only be unambiguously answered if thermodynamic analysis is combined with dynamic and structural studies.

One of the arguments in favor of the discrete approach is the direct computation of the equilibrium number of molecules involved in alternative states in several simulated water-like models. The fractions of molecules involved in the high-density structure and in the low-density structure at various temperatures and pressures have been computed for the ST2 [95], TIP4P/2005 [96] and mW [66, 97] models. While being well described by the two-state thermodynamics, the mW model does not exhibit liquid-liquid separation, behaving similar to the singularity-free scenario. We note that more accurate atomistic models of water are available for bulk properties [98], which have not yet been applied to this problem. In particular, the role of polarization is being increasingly recognized as having a significant influence on the properties of water [99, 100] and has not been considered so far.

The existence of a bimodal distribution of molecular configurations in real water is supported by X-ray photon correlation spectroscopy (XPCS), [101], and by an investigation of vibrational dynamics [102]. An unresolved theoretical problem is the microscopic nature of the phenomenological order parameter (the molecular fraction of conversion in the two-state thermodynamics) associated with the bimodal distribution in supercooled water. The concept of locally favored structures, developed by Tanaka et al. [51-54, 88], accounting for coupling between the orientational and translational local orders are promising steps in resolving this problem.

An unexplored area, both experimentally and computationally, is the kinetics of interconversion of the alternative structures. The chemical relaxation rate becomes slower upon cooling and thus may interplay with the rate of phase transformations. The rate interconversion of discrete molecular or supramolecular forms depends on the activation barrier. The existence of the activation barrier is a signature of interconversion. This barrier can be tuned both experimentally (catalysis, temperature) and computationally (by simulating intermediate states). The rate of interconversion can be obtained by measuring the relaxation of fluctuations of reaction coordinate by dynamic light scattering (see Supplemental Material, Section 8).

It is known that conserved and non-conserved order parameters may belong to different classes of universality in dynamics [103]. The reaction coordinate is a non-conserved order parameter that obeys the dynamics of relaxation independent of the wave number. Density and entropy are conserved quantities. They obey a diffusive relaxation with the rate proportional to the square of wave number. Therefore, experimental and simulation studies of the relaxation rate at different wave numbers could discriminate the nature of polyamorphism.

Another unresolved question is the relation between the developed phenomenology of discrete alternative states and a two-scale isotropic intermolecular potential, such as the Jagla potential [104-107] or, more generally, soft-repulsion potentials [108, 109] that generate a liquid-liquid transition in a single-component system. As pointed out by Vilaseca and Franzese [108], isotropic intermolecular potentials, due to the lack of directional interactions, provide a mechanism for fluid polyamorphism that is an alternative to the bonding in network-forming liquids, such as water. It seems that the entropy of the systems described by an isotropic intermolecular potential may not contain the term that is associated with the entropy of mixing of two discrete states. However, how can the molecular clustering observed in simulations of a Jagla-potential fluid [110] be interpreted? In the discrete lattice-gas model there is no distance-dependent intermolecular potential. The discrete lattice-gas model and continuous van der Waals model can be reconciled by a crossover procedure known as the "fine lattice discretization" [60]. How could this procedure affect the evolution of the shape of intermolecular potential? Ultimately, any peculiarities in the condensed-matter behavior are determined by details in interatomic and intermolecular interactions. Answers to the questions raised are highly desirable and require further investigation.

Finally, the microscopic nature, and even the existence, of polyamorphism can be elucidated by studying the phase transitions in binary solutions, which stem from polyamorphism predicted, but inaccessible, in the pure solvent [111-117]. Liquid-liquid transitions in binary solutions usually originate from essential nonideality of mixing. However, if a liquid-liquid transitions is found in an ideal solution, this transition must be stemming from the liquid-liquid transition of the pure solvent [116]. Therefore, a recent calorimetric study [117] of an ideal solution (hydrazinium trifluoroacetate in water) is probably the most direct evidence, obtained so far, for water's polyamorphism.

## IV. CONCLUSIONS

Fluid polyamorphism, or, more generally, "fluid polymorphism" if liquid crystals are included, is a surprisingly ubiquitous, yet poorly understood, phenomenon in condensed matter, either observed or predicted in a broad range of materials. We have developed a generic phenomenological approach, based on the Landau theory of phase transitions, to describe polyamorphism in a single-component fluid. It is completely independent of the underlying molecular origin of the phenomenon and sheds new light on the physical nature of polyamorphism.

We utilized the concept of thermodynamic equilibrium between two competing interconvertible states or molecular structures. The existence of two competing states in a single-component fluid may promote fluid polyamorphism either with or without phase separation depending on the symmetry of the order parameter. If the order parameter is a scalar, associated with the molecular fraction of conversion, the polyamorphism is accompanied by fluid phase separation. If the order parameter is a vector (the lambda transition in helium and an infinite-degree polymerization transition), the polyamorphism may be accompanied by phase separation only at the account for coupling of the vector order parameter with a scalar order parameter, thus causing tricriticality.

The two-state thermodynamics naturally unifies all the debated cases of fluid polyamorphism: with and without phase separation, from the "singularity-free" scenario to the "critical-point-free" one and qualitatively describes the thermodynamic anomalies typically observed in polyamorphic materials. We have discovered a remarkable peculiarity of liquid polyamorphism, which has not been reported previously in the literature, a singularity ("bird's beak") in the liquid-liquid coexistence curve when the critical point coincides with the liquid-vapor spinodal. This singularity is a generic feature, being associated with the common tangent of the liquid-liquid coexistence and vapor-liquid spinodal at the temperature-density and pressure-density planes.

The developed approach enables a global equation of state to be formulated that uniformly describes both vapor-liquid and liquid-liquid equilibria in single-component fluids, including the metastable and doubly metastable states under negative pressures. Further experimental and simulation studies of dynamic and structural properties are desirable to verify other predictions of this approach, such as shrinking the locus of density maximum in the critical-point-free scenario, and elucidate the microscopic foundation of the developed phenomenology.

Our work makes a paradigm shift from fluid polyamorphism defined as a relatively narrow phenomenon of liquid-liquid separation in a single-component fluid to a cross-disciplinary field that addresses a broad class of systems and phenomena with interconversion of alternative molecular or supramolecular states. The phenomenology developed in our work extends the original two-state model far beyond liquid-liquid separation in a single-component fluid and opens the way to construct global equations of state for various materials of physically different nature, polyamorphic or not, wherever molecular interconversion may take place.

## ACKNOWLEDGEMENTS

We acknowledge fruitful discussions with C. A. Angell, V. V. Brazhkin, S. V. Buldyrev, P. G. Debenedetti, M. E. Fisher, G. Franseze, J. Hrubý, P. Poole, V. N. Ryzhov, H. E. Stanley, and J. V. Sengers. K. Amman-Winkel made a valuable comment on the X-ray technique used in ref. [90]. M. A. Anisimov thanks Swinburne University of Technology, Australia for funding under the auspices of the Distinguished Visiting Researcher Scheme, the Institute of Multiscale Science and Technology (Labex iMUST) supported by the French ANR, and by the Russian Foundation for Basic Research (Grant No. 16-03-00895-a) during part of his sabbatical leave from the University of Maryland. The research of M. Duška at the University of Maryland was supported by the Young Scientist IAPWS Fellowship Project "Towards an IAPWS Guideline for the Thermodynamic Properties of Supercooled Heavy Water".

# Thermodynamics of Fluid Polyamorphism

Mikhail A. Anisimov[1,2*], Michal Duška[1,3], Frédéric Caupin[4], Lauren E. Amrhein[1], Amanda Rosenbaum[1], and Richard J. Sadus[5]

[1]*Department of Chemical & Biomolecular Engineering and Institute for Physical Science & Technology, University of Maryland, College Park, U.S.A.*
[2]*Oil and Gas Research Institute of the Russian Academy of Sciences, Moscow, 117333, Russia*
[3]*Institute of Thermomechanics, Academy of Sciences of the Czech Republic, 182 00 Prague 8, Czech Republic*
[4]*Université de Lyon, Université Claude Bernard Lyon 1, CNRS, Institut Lumière Matière, F-69622 Villeurbanne, France*
[5]*Centre for Molecular Simulation, Swinburne University of Technology, Hawthorn, Victoria 3122, Australia*

*To whom correspondence should be addressed. Email: anisimov@umd.edu.

## 1. Lattice-gas model

The lattice-gas model was first introduced by Frenkel in 1932 [1]. In 1952 Yang and Lee [2] showed that the lattice-gas model, which is the simplest model for the vapor-liquid transition, is mathematically equivalent to the Ising model that describes a phase transition between paramagnetic and anisotropic ferromagnetic or antiferromagnetic states. The Ising/lattice-gas model is also used to describe solid-solid phase separation or order-disorder transitions in binary alloys as well as liquid-liquid phase separation in binary fluids. The model plays a special role in the physics of condensed matter because it can be applied to very different systems and phenomena, thereby bridging the gap between the physics of fluids and solid-state physics [3-5]. The volume of the system is divided into cells of molecular size $l_\text{o}$. These cells are arranged in a lattice with the coordination number $z$ ($z$ = 6 for a simple cubic lattice). In its simplest version each cell in the lattice is either empty or occupied by only one molecule. The molecular density in the lattice gas model is dimensionless, being defined as $\rho = l_\text{o}^3 \left( N/V \right)$, where $N$ is the number of occupied cells. The nearest-neighboring molecules interact by short-ranged attractive forces. Empty cells do not interact with their neighbors.

The exact analytical solution for the three-dimensional lattice gas has not been reported, although accurate numerical solutions are available. The simplest approximate analytical solution can be obtained in the mean-field approximation, with the multi-body attraction energy being represented by a single interaction parameter $\varepsilon$. This approximation is equivalent to accounting for attraction in the van der Waals fluid via the constant $a$. Similarly, the molecular volume of the lattice gas ($l_o^3$) is equivalent to the van der Waals co-volume parameter ($b$).

The equation of state for lattice gas in the mean-field approximation reads

$$p = -k_B T \ln(1-\rho) - \varepsilon \rho^2, \tag{S1}$$

where $p$, $T$, and $k_B$ are the pressure, temperature and Boltzmann constant, respectively. In Eq (S1), the pressure is in units of energy and the density is dimensionless. The chemical potential $\mu$ of the lattice gas and the density of the Helmholtz energy $\rho F = \rho \mu - p$ in the mean-field approximation are

$$\mu = \left[\frac{\partial(\rho F)}{\partial \rho}\right]_T = k_B T \ln\frac{\rho}{1-\rho} - 2\varepsilon\rho + \varepsilon + \varphi(T) \tag{S2}$$

and

$$\rho F = \rho\mu - p = k_B T \rho \ln\rho + k_B T(1-\rho)\ln(1-\rho) + \varepsilon\rho(1-\rho) + \rho\varphi(T). \tag{S3}$$

The function $\varphi(T)$ represents the Helmholtz energy of ideal gas per molecule that depends on temperature only. This function does not affect the phase equilibrium but is needed to calculate the heat capacity.

After subtracting the terms linear in $\rho$, the Helmholtz energy obtained from Eq. (S3) is symmetric with respect to $\rho$. This means that, unlike that in the essentially asymmetric van der Waals fluid, the condition for fluid-phase equilibrium (binodal) for the lattice gas can be found analytically from

$$\mu_{\text{cxc}} - \varphi(T) = k_\text{B} T \ln \frac{\rho}{1-\rho} - 2\varepsilon\rho + \varepsilon = 0 . \tag{S4}$$

The vapor-liquid spinodal

$$\left(\frac{\partial \mu}{\partial \rho}\right)_T = \frac{k_\text{B} T}{\rho(1-\rho)} - 2\varepsilon = 0 \tag{S5}$$

The critical-point parameters are $\rho_\text{c} = 1/2$, $T_\text{c} = \varepsilon/2k_\text{B}$, $p_\text{c} = k_\text{B} T_\text{c} \left[ -\ln(1/2) - 1/2 \right] \cong 0.193 k_\text{B} T_\text{c}$.

In the reduced form ($\hat{T} = T/T_\text{c}$, $\hat{p} = p_\text{c}$) the lattice-gas equation of state reads

$$\hat{p} = \frac{\hat{T} \ln(1-\rho) + 2\rho^2}{\ln\frac{1}{2} + \frac{1}{2}} \cong -5.18 \left[ \hat{T} \ln(1-\rho) + 2\rho^2 \right]. \tag{S6}$$

The critical-point parameters are $\rho_\text{c} = 1/2$, $\hat{T}_\text{c} = 1$, and $\hat{p}_\text{c} = 1$.

The equation for the liquid-vapor spinodal ("sp") reads

$$\hat{T}_{\text{sp}} = 4\rho(1-\rho) . \tag{S7}$$

The equation for the liquid-vapor coexistence ("cxc") reads

$$\frac{1}{\hat{T}_{\text{cxc}}} = \frac{\frac{1}{2} \ln \frac{\rho}{1-\rho}}{2\rho - 1} . \tag{S8}$$

The lattice-gas model mathematically equivalent to the Ising model of an incompressible anisotropic ferromagnetic material. In zero magnetic field, the magnetization, $M$, spontaneously emerges at a certain temperature i.e., the "Curie temperature". Introducing $M = 2\rho - 1$ and using the mathematical fact that $\operatorname{arctanh}(M) = \frac{1}{2} \ln \frac{1+M}{1-M}$, we obtain $\frac{M}{\hat{T}_{\text{cxc}}} = \frac{1}{2} \ln \frac{1+M}{1-M}$. Hence,

$$M = \tanh \frac{M}{\hat{T}_{\text{cxc}}} . \tag{S9}$$

Equation (9) is commonly used for the Ising model, describing the spontaneous magnetization below the Curie point of a ferromagnetic-paramagnetic transition.

Figures S1 and S2 demonstrate the phase behavior and properties of the lattice-gas model.

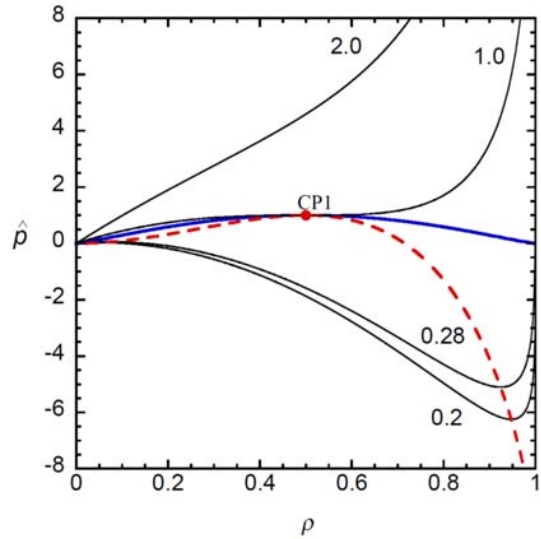

**Fig. S1.** Selected isotherms and liquid-vapor coexistence of the lattice-gas model. Sold blue is binodal. Dashed red is spinodal. Red dot is the liquid-vapor critical point.

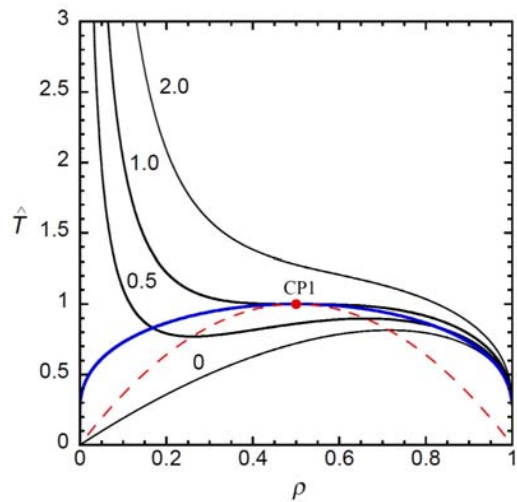

**Fig. 2.** Selected isobars and liquid-vapor coexistence of the lattice-gas model. Solid blue is binodal. Dashed red is spinodal. Red dot is the liquid-vapor critical point.

## 2. "Fine lattice discretization": crossover between lattice gas and van der Waals fluid

Moghaddam et al. [6] developed a procedure, "fine lattice discretization", describing crossover between two limits: the discrete, lattice-gas model and the continuous, van der Waals model. Helmholtz energy per unit volume and chemical potential for lattice gas:

$$\rho F = \rho\mu - p = kT\rho\ln\rho + kT(1-\rho)\ln(1-\rho) + \varepsilon\rho(1-\rho) + \rho\varphi(T). \tag{S10}$$

$$\mu = \left[\frac{\partial(\rho F)}{\partial \rho}\right]_T = kT\ln\frac{\rho}{1-\rho} - 2\varepsilon\rho + \varphi(T). \tag{S11}$$

$$\mu_{cxc} - \mu_0(T) = kT\ln\frac{\rho}{1-\rho} - 2\varepsilon\rho = 0. \tag{S12}$$

For the van der Waals fluid, the density of the Helmholtz energy is

$$\rho F = kT\rho\ln\frac{\rho}{1-\rho} - \varepsilon\rho^2 + \rho\varphi(T) \tag{S13}$$

In Eq. (S13) we use the following rescaling of the physical density and the van der Waals parameter $a$: $\rho \to b\rho$ and $a \to \varepsilon = a/b$.

$$\mu = \left[\frac{\partial(\rho F)}{\partial \rho}\right]_T = kT\ln\frac{\rho}{1-\rho} + \frac{kT}{1-\rho} - 2\varepsilon\rho + \varphi(T). \tag{S14}$$

$$\rho_c = \frac{1}{3}, \; T_c = \frac{8\varepsilon}{27kb}. \tag{S15}$$

Fine lattice discretization model:

$$\frac{\mu - \mu_0(T)}{kT} = \ln\frac{\rho}{1+\rho(1/\zeta-1)} + \zeta\ln\left(1+\frac{\rho}{\zeta(1-\rho)}\right) - \frac{2a\rho}{kT}. \tag{S16}$$

Discretization parameter $\zeta = l/l_0$ ($l$ molecular diameter, $l_0$ lattice spacing)
$\zeta = 1$ → lattice gas
$\zeta \to \infty$ → van der Waals fluid

$$\rho_c = \frac{1}{2}\psi_1(\zeta), \; T_c = \frac{\varepsilon}{2k}\psi_2(\zeta).$$

The critical density functions of the discretization parameter.

3. **Fluid polyamorphism caused by thermodynamic equilibrium between interconvertible states A and B: an alternative choice of the equilibrium constant as a function of temperature and pressure**

Consider an alternative choice of $\ln K(\hat{T}, \hat{p})$ for the "reaction" $A \rightleftarrows B$ to be a linear function of temperature and pressure:

$$-\ln K(\hat{T}, \hat{p}) = \lambda'(\hat{T} + \alpha'\hat{p} + \beta'), \tag{S17}$$

$$\hat{G}_{BA}(\hat{T}, \hat{p}) = -\hat{T} \ln K(\hat{T}, \hat{p}) = \lambda'(\hat{T}^2 + \alpha\hat{p}\hat{T} + \beta\hat{T}). \tag{S18}$$

This choice corresponds to the changes in both reaction energy ($\hat{U}_{BA} = \lambda'\hat{T}^2$) and the volume ($\hat{V}_{BA} = \alpha'\hat{T}$) to vanish at zero temperature. The line $\hat{T} + \alpha'\hat{p} + \beta' = 0$ is the locus of liquid-liquid transitions. The $p-T$ and $T-\rho$ phase diagrams for a specific case with $\lambda' = 10$, $\alpha' = 0.02$, and $\beta' = -0.3$ is presented in Figs. S3 – S7. We observed that when the volume change of reaction vanishes at zero temperature, the density difference between the coexisting liquid phases also vanishes.

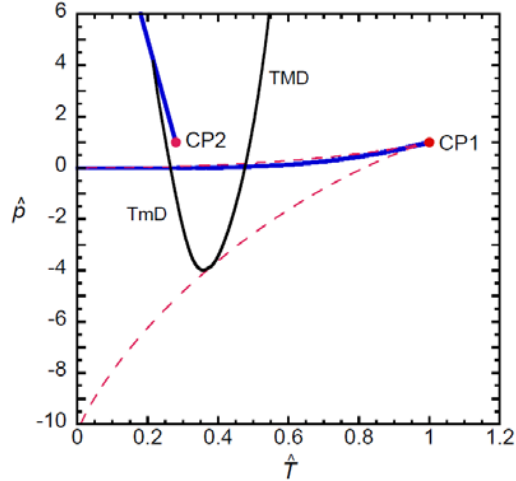

**Fig. S3.** *P-T* diagram of lattice gas with the second (liquid-liquid) transition. Two blue curves are vapor-liquid and liquid-liquid transitions terminated by the critical points CP1 and CP2 (shown by red dots), respectively. Solid black are the loci of minimum density (TmD) and maximum density (TMD). Dashed reds are two branches of the vapor-liquid spinodal. Thermodynamically, liquid can exist between the vapor-liquid coexistence and the low branch of spinodal as a metastable state.

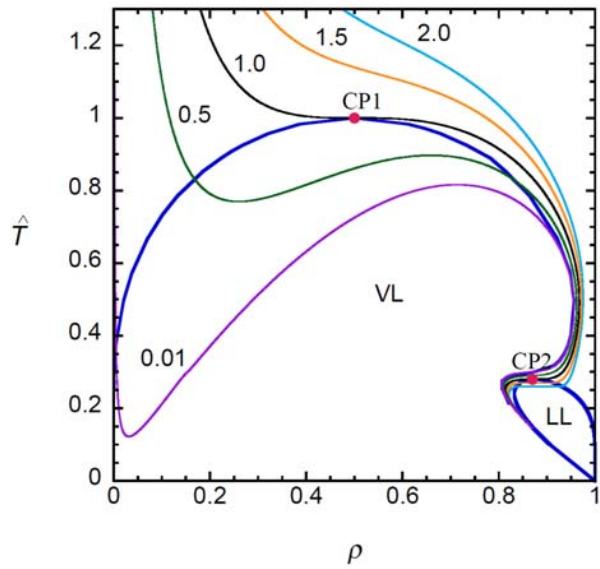

**Fig. S4.** Selected isobars, liquid-vapor (LV) and liquid-liquid (LL) coexistence (both shown by thick blue curves) of the lattice-gas model with "chemical reaction". Solid blue curves are the binodals. Red dots are the liquid-gas and liquid-liquid critical points.

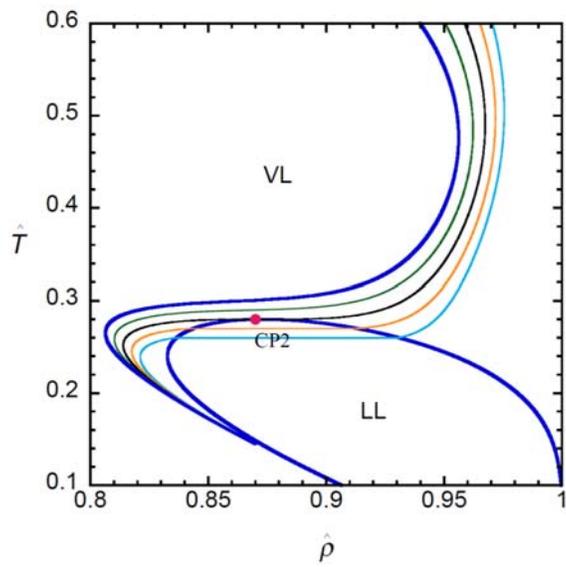

**Fig. S5.** Selected isobars in the vicinity of the liquid-liquid coexistence (shown by thick blue) of the lattice-gas model with "chemical reaction".

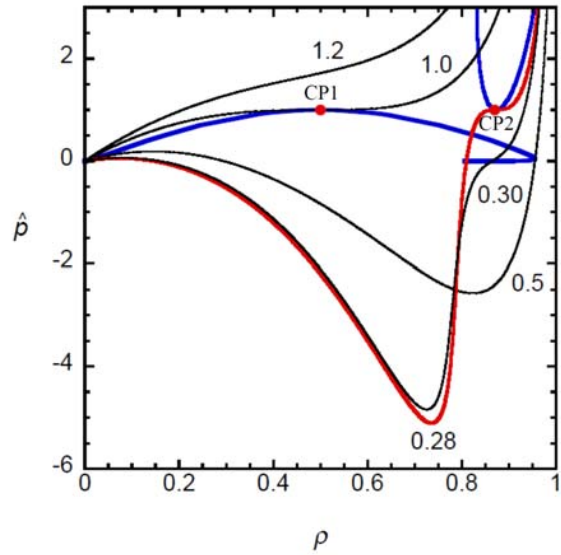

**Fig. S6.** Selected isotherms, liquid-vapor and liquid-liquid coexistence (both shown by thick blue curves) of the lattice-gas model with "chemical reaction". Red curve is the critical isotherm of liquid-liquid coexistence. Red dots are the liquid-gas and liquid-liquid critical points.

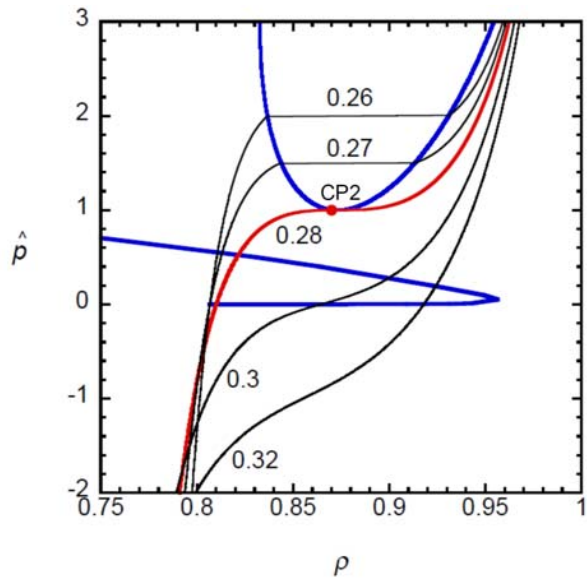

**Fig. S7.** Selected isotherms in the vicinity of the liquid-liquid coexistence (shown by thick blue) of the lattice-gas model with "chemical reaction". Red curve is the critical isotherm of liquid-liquid coexistence. Red dot is the liquid-liquid critical point.

The fraction of state B as a function of temperature for selected isobars is presented in Fig. S8.

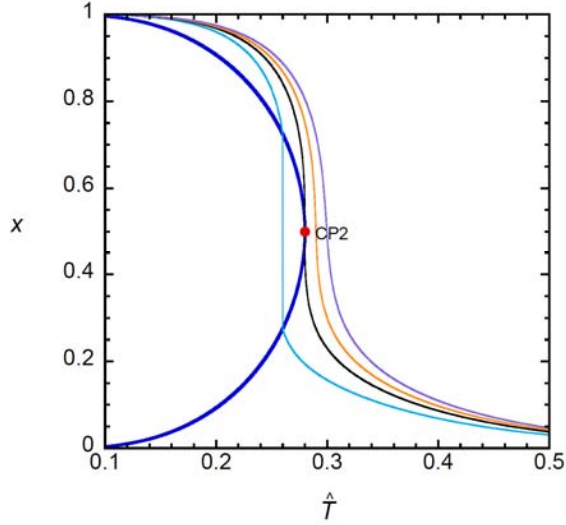

**Fig. S8.** Fraction of state B as a function of temperature for selected isobars (from left to right: $\hat{p}$ = 2.0, 1.0, 0.5, and 0.01) and along the liquid-liquid coexistence ($\ln K(T, p) = 0$, thick blue). Red dot is the liquid-liquid critical point.

4. **Fluid polyamorphism (the van der Waals equation as a choice for state A): tuning the location of the liquid-liquid critical point**

If the van der Waals model is chosen for state A then the phase diagrams predicted by two-state thermodynamics turns out to be essentially similar to those of the lattice-gas model for state A. The evolution of the extrema loci upon tuning the location of the liquid-liquid critical point is demonstrated in Figure S9 from (A) (singularity-free scenario) to (D) (critical point-free scenario). The disappearance of the maximum density locus in the case where the critical point coincides with the vapor-liquid spinodal point is demonstrated in Figure S10 by zooming (C). Figures S9 A to D show the extrema patterns similar to those in Figure 4 of the main text. The parameters of the liquid-liquid transition/Widom line, as defined by Eq. (15) of the main text, in the van der Waals case are $\lambda/kT_{c1} = 0.5$, $\alpha/\rho_{c1}T_{c1} = -0.05$, $\beta/k = -2$.

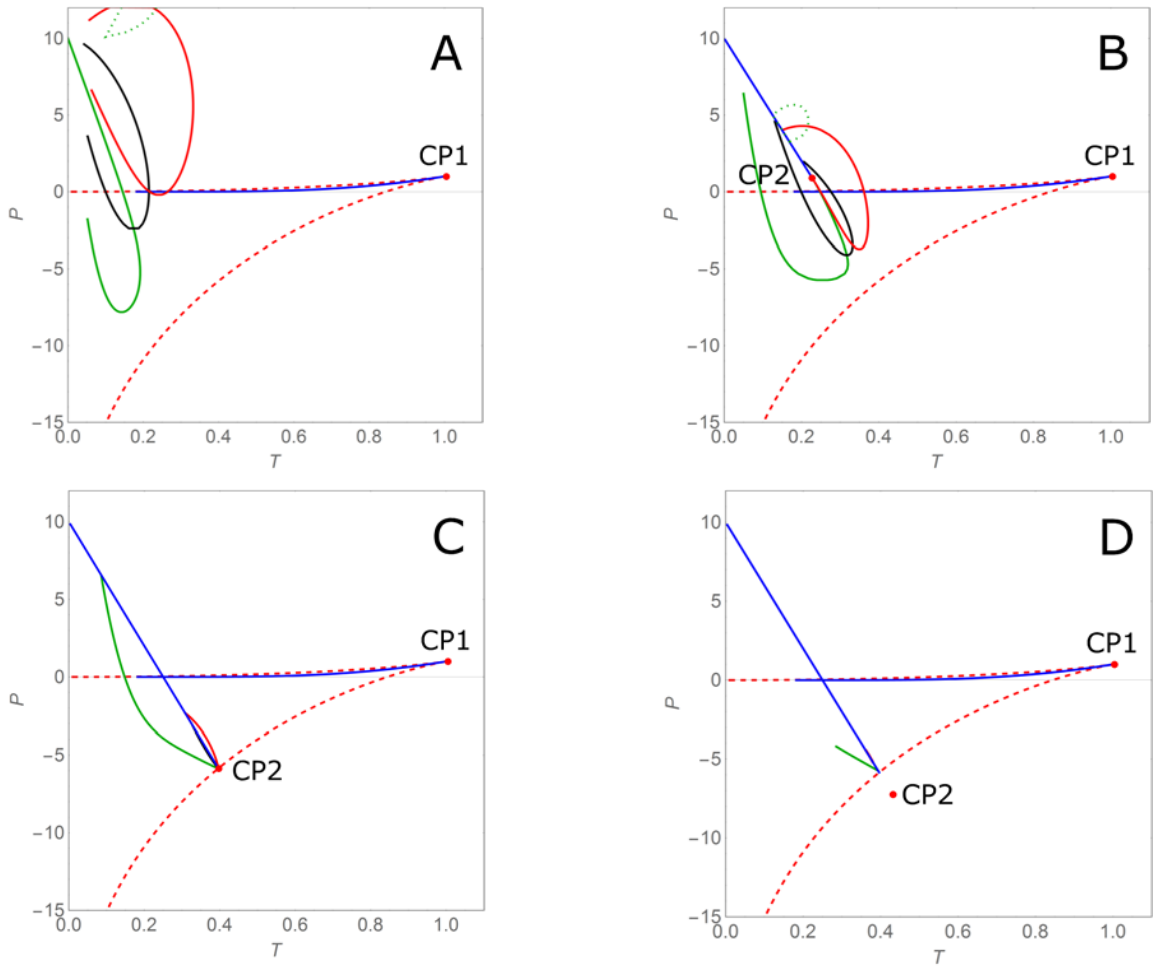

**Figure S9.** Evolution of the pattern of the extrema loci upon tuning the location of the liquid-liquid critical point, when state A is described with the van der Waals equation of state. Blue are liquid-vapor and liquid-liquid equilibrium lines, black is the density maximum or minimum; red is the isothermal compressibility maximum or minimum along isobars, green is the isobaric heat capacity maximum or minimum along isotherms; dashed green shows additional (shallow) extrema of the heat capacity unrelated to the liquid-liquid transition; dashed red are two branched of the liquid-vapor spinodal; red dot is the liquid-liquid critical point. (A) a singularity-free scenario – the critical point is at zero temperature; (B) a "regular" scenario – the critical point is at a positive pressure; (C) the critical point coincides with the absolute stability limit of the liquid state; (D) a critical point-free scenario – the "virtual" critical point is located in the unstable region.

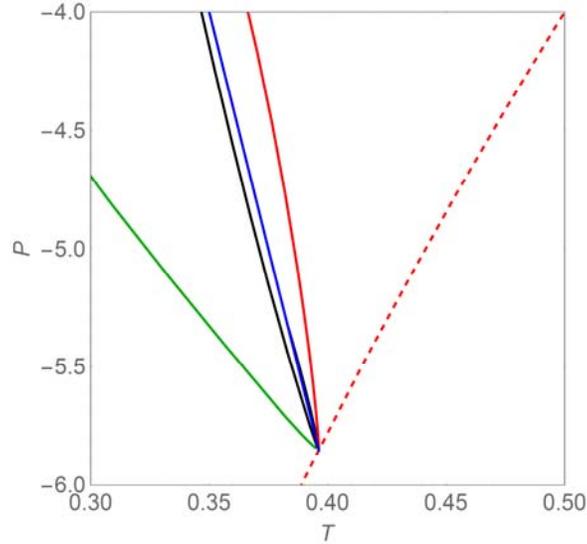

**Fig. S10**. Zooming the area near the liquid-liquid critical point shown in Figure S9C, which demonstrates the disappearance of the maximum density locus in the case where the critical coincides with the vapor-liquid spinodal point. The minimum density locus (to the left of the liquid-liquid coexistence) still exists.

5. **Fluid polyamorphism: a singularity of the liquid-liquid coexistence ("bird's beak" ), the two-state model with van der Waals equation as the choice for state A**

Fig. S11 demonstrates the gradual development of a singularity of the liquid-liquid coexistence when the metastable (with respect to vapor) liquid-liquid critical point approaches the absolute stability limit of liquid. A completely developed singularity ("bird's beak") is observed when the liquid-liquid critical point coincides with the liquid-vapor spinodal as shown in Fig. S11C and D. The character of the singularity for this (van der Waals) choice of state A is essentially similar to that observed for the lattice-gas choice of state A, confirming that the phenomenon is generic (compare with Fig. 5b of the main manuscript).

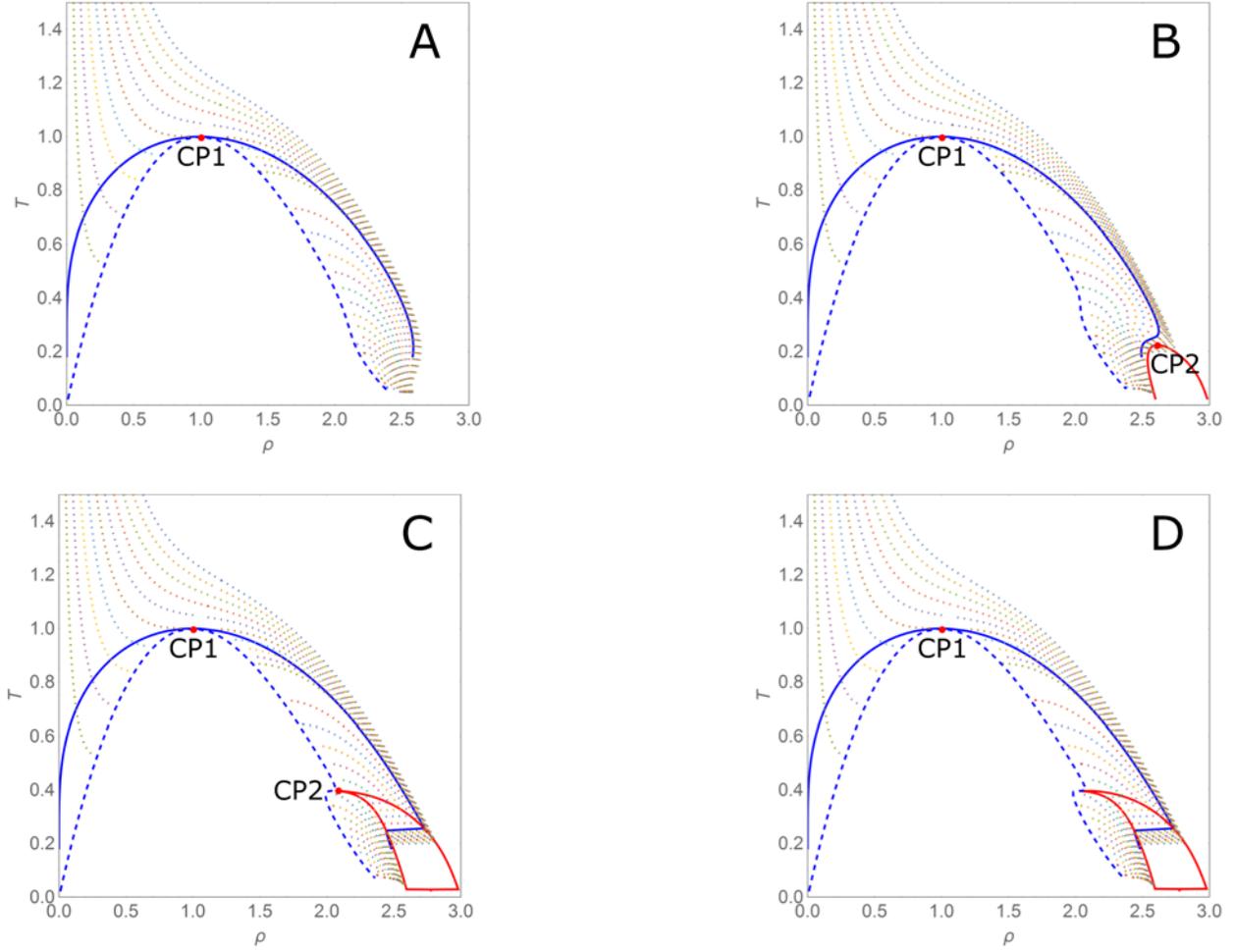

**Figure S11.** A singularity ("bird's beak") in the liquid-liquid coexistence curve, which develops when the critical point approaches with the liquid-vapor spinodal (the van der Waals choice for state A). The four $T-\rho$ diagrams correspond to the four cases shown in Fig. S9. Blue curve is the vapor-liquid coexistence; red is the liquid-liquid coexistence; dashed blue is the liquid-vapor spinodal; multicolor curves are selected isobars.

## 6. Fluid polyamorphism: phase separation near the tricritical point of a polymerization transition

The phase separation of a polymer solution into the pure monomer solvent and concentrated polymer solution in the limit of the infinite degree of polymerization is obtained from the condition

$$-kT\ln(1-x) - kTx - \frac{1}{2}\Theta x^2 = 0. \qquad (S19)$$

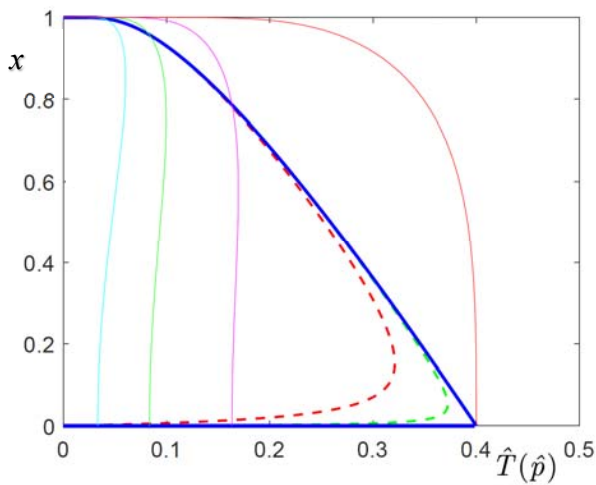

**Fig. S12.** Two solutions of Eq. (19) for the fraction of polymerized molecules ($x$) below the tricritical (theta) point in the limit of infinite degree of polymerization. Blue is the liquid-liquid coexistence. Dashed curves correspond to finite degree of polymerization (red $N=10$, green $N=30$). Multicolor curves are selected isobars.

7. **Transient equation of state**

   In practice, one can expect the situation when the characteristic time of reaction and the time of experiment are comparable. This is the case of strong coupling between thermodynamics and kinetics. This area is most promising for new discoveries. For example, isomerization of a hydrocarbon, such as interconversion of *n*-butane and isobutane, is within the framework of the singularity-free scenario (Figure S9a). Without a catalyst, isomerization of butane is extremely slow. In the limit of zero conversion rate the system thermodynamically behaves as a two-component mixture with corresponding stability criteria. This is why one can distill hydrocarbon isomers, store them separately, mix and apply appropriate binary-solution model to describe their thermodynamic properties typical for a binary fluid, e.g., the lines of critical and triple points. Significantly, the isothermal compressibility and isobaric heat capacity do not demonstrate a strong, van der Waals-like divergence at the critical point of a fluid mixture.

   Even in the presence of catalysts such as aluminum halides, the time of reaching the full equilibrium conversion may take days [7]. However, some recently reported [8] catalytic techniques can increase the rate of isomerization by many orders of magnitude, up to $1.5 \cdot 10^9$ s$^{-1}$, thus making the mixture of two isomers thermodynamically equivalent to a single-component fluid (if the observation time greater than nanosecond), with the isomer ratio being the function of pressure and temperature. In this system, one cannot separate A from B by a slow separation technique.

If an equation of state of isomerizing butane is to be transient then it should contain an additional parameter (let it be notated as $\zeta$), the product of the reaction rate and the characteristic process time. It will have two thermodynamic limits, namely single-component fluid ($\zeta \to \infty$) and a binary solution $(\zeta \to 0)$. Between these two limits, the equation describes a nonequilibrium state and is controlled by the ratio of the two characteristic times. The transient equation of state will give a snapshot of the nonequilibrium reacting fluid at any stage of reaction. Significantly, the parameter $\zeta$ is a function of temperature and depends on a particular catalyst.

In the limit, $\zeta \to \infty$, the equation of state obeys the stability criteria of a single-component system and should demonstrate a unique critical point with strongly divergent isothermal compressibility and isobaric heat capacity at the critical point. Moreover, the competition between the isomers having different properties in their pure states (3-5 % difference for butanes) will generate the thermodynamic anomalies and lines of extrema expected for the singularity-free scenario as demonstrated in Figure S4a.

## 8. Relaxation of fluctuations in chemically reacting fluids

It is well known that fluctuations in fluids can be probed by light scattering [9, 10]. However, light-scattering studies of chemically reacting fluid mixtures have not received wide implementation. Light scattering is primarily suited for studying kinetics of relatively fast chemical reactions: the reaction rate is to be larger than the diffusion relaxation rate [11-13]. In principle, dynamic light scattering [9] is best suitable for studying kinetics of chemical reactions in fluids. The beauty of this method that the system could be in equilibrium but the fluctuations randomly emerge and disappear ("relax") in accordance with non-equilibrium thermodynamics [14]. Fluctuations of the reaction coordinate are coupled with fluctuations of density and concentration and can be probed accordingly. The characteristic rates of chemical reaction to be detected by dynamic light scattering are in the range from $10^8$ s$^{-1}$ to 0.1 s$^{-1}$.

The main issue here is that the decay rate of the fluctuations of reaction coordinate around reaction equilibrium, unlike diffusive relaxation of concentration fluctuations, does not depend on the wave number. However, a coupling between these two dynamic modes may change the story. One can expect a fundamental difference in the spectrum of fluctuations depending on whether the wavelength of the fluctuations is smaller or larger than the penetration length of the chemical reaction [14

Small-angle neutron scattering experiments, combined with polarimetry measurements, indicate that in isobutyric acid and isobutyric acid-rich aqueous solutions the polyethylene glycol (PEG) polymer chains (coils) at 55° C coexist with stiff rods (helices) at high molecular weights of PEO but at a low molecular weight the interconversion is shifted to the polymer rods [15]. The SANS data are consistent with the length of the rods of about 20 nm and diameter ~3 nm. It was also shown that the formation of helices by PEG in isobutyric acid requires the presence of a trace amount of water, even if PEG does not form helices in water [16], thus suggesting that water serves as a catalyst for the helix-coil reaction. In addition, the critical concentration fluctuations near the liquid-liquid critical point of isobutyric acid-water solution interplay with the coil-helix interconversion such that the conversion is suppressed in the critical region [17].

Depolarized dynamic light scattering, similar to that used to study self-assembly of cromolyn disc-like molecules into rods [18] can be used for measuring the rate of helix-coil interconversion in PEG-isobutyric solutions. The existence of the anisotropic helix rods will generate strong depolarized light scattering. In particular, it is needed to measure the wave-number dependence of the polarized and depolarized light scattering in order to investigate expected coupling between diffusive relaxation of concentration fluctuations and non-diffusive relaxation of fluctuations of reaction coordinate. Tuning the rate of the helix-coil conversion can be made by changing trace amounts of water in isobutyric acid.

Self-assembly provides a good example of the extension of the two-state thermodynamics to binary solutions. Reversible self-assembly of cromolyn disc-like molecules into rods promote liquid-liquid separation in cromolyn aqueous solutions [18, 19]. Rods and discs are unequally distributed in the coexisting phases. While the dilute phase is isotropic, the phase enriched with rods is a nematic liquid crystal. The tensor character of the orientational order parameter, coupled with concentration (scalar), makes the phase transition to be first-order.